\documentclass{article}

\usepackage{verbatim}
\usepackage{graphicx}
\usepackage{natbib}
\usepackage{bm}
\usepackage{amssymb}
\usepackage{amsmath}
\usepackage{authblk}

\DeclareMathOperator*{\argmax}{arg\,max}

\title{Modeling neural activity at the ensemble level}

\author[1,2]{Joaqu\'{i}n Rapela\footnote{rapela@ucsd.edu}}
\author[3]{Mark Kostuk}
\author[4]{Peter F. Rowat}
\author[1]{Tim Mullen}
\author[5]{\\Edward F. Chang}
\author[6]{Kristofer Bouchard}
\affil[1]{Swartz Center for Computational Neuroscience, UCSD}
\affil[2]{Instituto de Investigación en Luz Ambiente y Visión, UNT and
CONICET, Argentina}
\affil[3]{Department of Physics, UCSD}
\affil[4]{Institute for Neural Computation, UCSD}
\affil[5]{Department of Neurological Surgery, UCSF}
\affil[6]{Computational Research Division, LBNL}
\date{}

\begin{document}

\maketitle
\tableofcontents

\pagebreak

\abstract{

Here we demonstrate that the activity of neural ensembles can be quantitatively
modeled. We first show that an ensemble dynamical model (EDM) accurately
approximates the distribution of voltages and average firing rate per neuron of
a population of simulated integrate-and-fire neurons. EDMs are high-dimensional
nonlinear dynamical models. To faciliate the estimation of their parameters we
present a dimensionality reduction method and study its performance with
simulated data.  We then introduce and evaluate a maximum-likelihood method to
estimate connectivity parameters in networks of EDMS. Finally, we show that
this model an methods accurately approximate the high-gamma power evoked by
pure tones in the auditory cortex of rodents. Overall, this article
demonstrates that quantitatively modeling brain activity at the ensemble level
is indeed possible, and opens the way to understanding the computations
performed by neural ensembles, which could revolutionarize our understanding of
brain function.

}

\pagebreak

\section{Introduction}

If we observe a fluid at the molecular level we see random motions, but if we
look at it macroscopically we may see a smooth flow. An intriguing possibility
is that by analyzing brain activity at a macroscopic level, i.e., at the level
of neural ensembles, we may discover patterns not apparent at the single-neuron
level, that are as useful as velocity or temperature are to understand, and
predict, the motion of fluids. 

Technology frequently drives science. For instance, thanks to the development
of microelectordes in the 1930's, we now know with exquiste detail computations
performed by single neurons. We are now experiencing a dramatic increase in our
capacity to monitor the activity of larger and larger populations of neurons
with higher and higher spatial and temporal resolution.  These new ensemble
recordings may soon allow us to uncover crucial computations performed by
neural ensembles.

Here we present results of developing and evaluating a mathematical model and
estimation methods to characterize the activity of ensembles of neurons from
electrophysiological data.

Section~\ref{sec:buildingEDMs} reports the evaluation of an ensemble dyanmical
model (EDMs). And EDMs is a high-dimensional nonlinear dynamical models. To
estimate its parameters it is convenient to reduce the number of parameters. We
escribe a dimensionality reduction method for EDMs in
Section~\ref{sec:reducingDimOfEDMs}. Section~\ref{sec:estimatingParamsOfEDMs}
presents a maximum-likelihood method to estimate parameters of EDMs. Finally,
in Section~\ref{sec:modelingRodentsRecordingsWithEDMs} we show that EDMs can
accurately approximate high-gamma electroencephalographic activity evoked by
pure tones in the auditory cortex of rodents.

\pagebreak

\section{Building EDMs}
\label{sec:buildingEDMs}

We wanted to learn how to build Ensemble Density Models (EDMs), dynamical
models of the state variables (e.g., trans-membrane potential, time since
last spike, etc.) of a population of identical neurons, starting from a
dynamical model of the state variables of a single neuron. For Integrate and
Fire (IF) model of single neurons, an EDM should provide the ensemble
probability density function (pdf) $\rho(\upsilon, t)$, from which to compute
the probability of finding a neuron in the ensemble with with a given
trans-membrane voltage $\upsilon$ at time $t$ (i.e., $P(\upsilon,
t)=\rho(\upsilon, t) dt$).  It should also provide the average firing rate
per neuron in the population, $r(t)$. To construct EDMs we chose the
methodology described in \citet{omurtagEtAl00}.

To evaluate the EDM we compared its outputs ($\rho(\upsilon, t)$ and $r(t)$)
with those derived from the direct simulation of a population of 9,000 IF
neurons. The value of the density function $\rho(\upsilon, t)$ derived from
the direct simulation was the proportion of IF neurons having a voltage
$\upsilon$ at time $t$, and the average firing rate per neuron $r(t)$ derived
from the the direct simulation was the proportion of cells in the population
at time $t$ with voltage at threshold.

An EDM is driven by an external excitatory current and modulated by an
external inhibitory current.  In addition, every cell in the EDM receives
inputs from $G$ other neurons in the population. A fraction $f$ of these $G$
inputs is excitatory, and the remainder are inhibitory.  These
intra-population inputs act as feedback mechanisms to the EDM.  The
mathematical representation of an EDM for a population of IF neurons is given
in Equations~\ref{eq:rhoDiffEq}--\ref{eq:J}, modified from Equations 26,
39 and 47 in \citet{omurtagEtAl00}. External excitatory and inhibitory
currents appears as $\sigma^0_e(t)$ and $\sigma^0_i(t)$, respectively, in
Equation~\ref{eq:J}, and excitatory and inhibitory feedback are given by the
terms $Gfr(t)$ and $G(1-f)r(t)$, respectively, in Equation~\ref{eq:J}.


\begin{eqnarray}
\frac{\partial\rho}{\partial t}(\upsilon, t)&=&-\frac{\partial J}{\partial\upsilon}(\upsilon,t)\label{eq:rhoDiffEq}\\
r(t)&=&J(\upsilon=1, t)\label{eq:r}\\
J(\upsilon,t)&=&-\gamma\upsilon\rho(\upsilon,t) \nonumber\\
& & +[\sigma^0_e(t)+Gfr(t)]\int_{\upsilon-h}^{\upsilon}\rho(\upsilon',t)d\upsilon'\nonumber\\ 
& & -[\sigma^0_i(t)+G(1-f)r(t)]\int_\upsilon^{\upsilon/(1-\kappa)}\rho(\upsilon',t)d\upsilon'
\label{eq:J}
\end{eqnarray}

We first verified that for a single population the outputs of the EDM matched
those of the direct simulations (Section~\ref{sec:onePopulation}). We next
built a network of excitatory and inhibitory populations, and again compared
the outputs of the EDM and those of the direct stimulation
(Section~\ref{sec:networkOfPopulations}).

\subsection{One Population}
\label{sec:onePopulation}

\subsubsection{One Population of Independent Neurons}

\begin{figure}
\begin{center}
\includegraphics[width=4in]{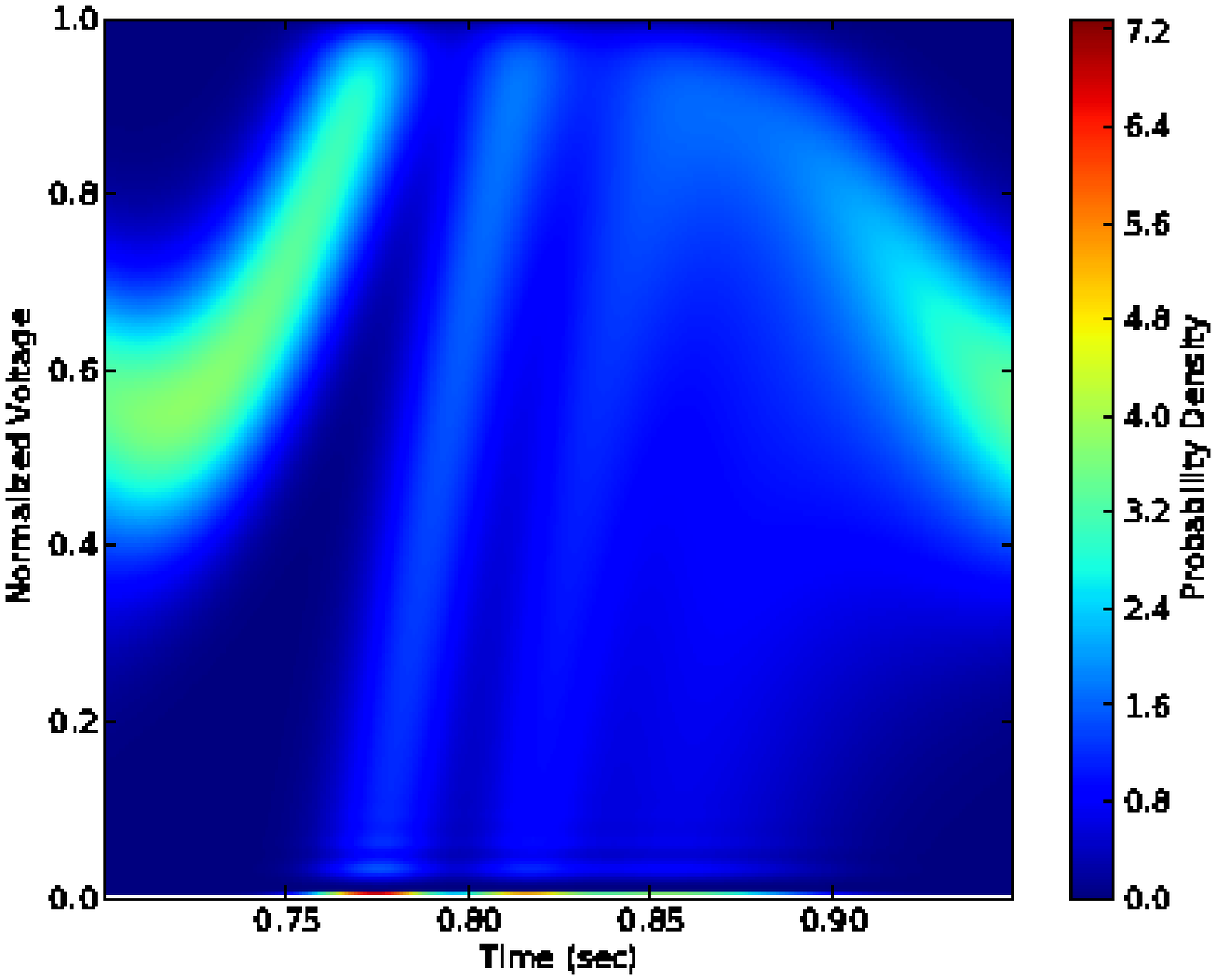}
\includegraphics[width=4in]{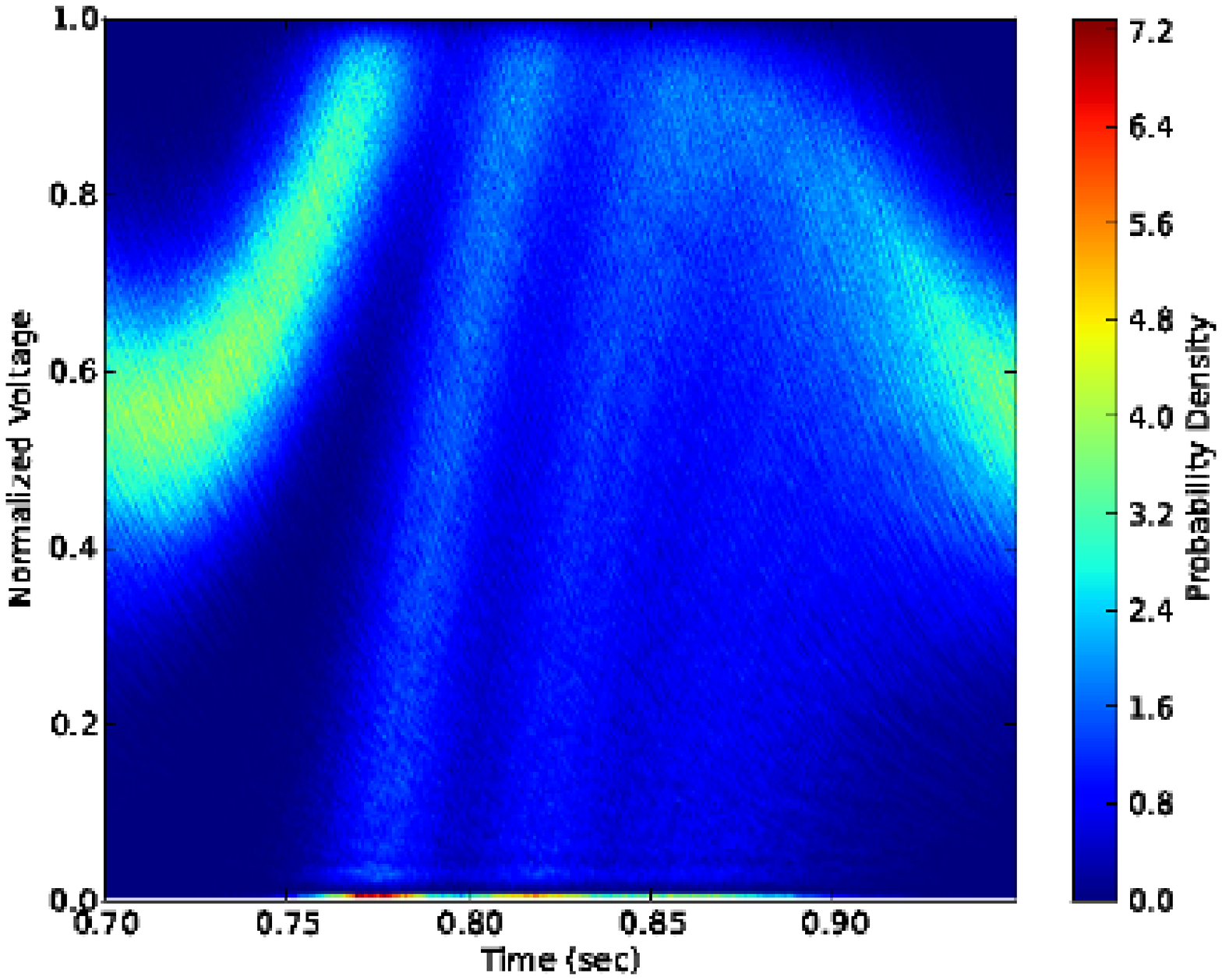}
\end{center}

\caption{Ensemble pdf, $\rho(\upsilon,t)$, for a population of IF neurons in
response to a sinusoidal input, calculated by integrating the differential
equation of an EDM (Equation~\ref{eq:rhoDiffEq}, top panel) and approximated
by direct simulation of a population of 9,000 IF neurons (bottom panel).
Neurons in this populations were independent of each other (i.e., they
had no feedback; G=0 in Equation~\ref{eq:J}).}

\label{fig:rhosSinusoidalRNFNI}
\end{figure}

The top panel in Figure~\ref{fig:rhosSinusoidalRNFNI} shows the ensemble
probability density function, $\rho(\upsilon,t)$, calculated by integrating the
differential equation of an EDM, Equation~\ref{eq:rhoDiffEq}. The bottom
panel shows and approximation to this pdf obtained from the histogram of
voltages of a direct simulation of a population of 9,000 IF neurons. Note the
large similarity of the pdfs at all time points.

\begin{figure}
\begin{center}
\includegraphics[width=5in]{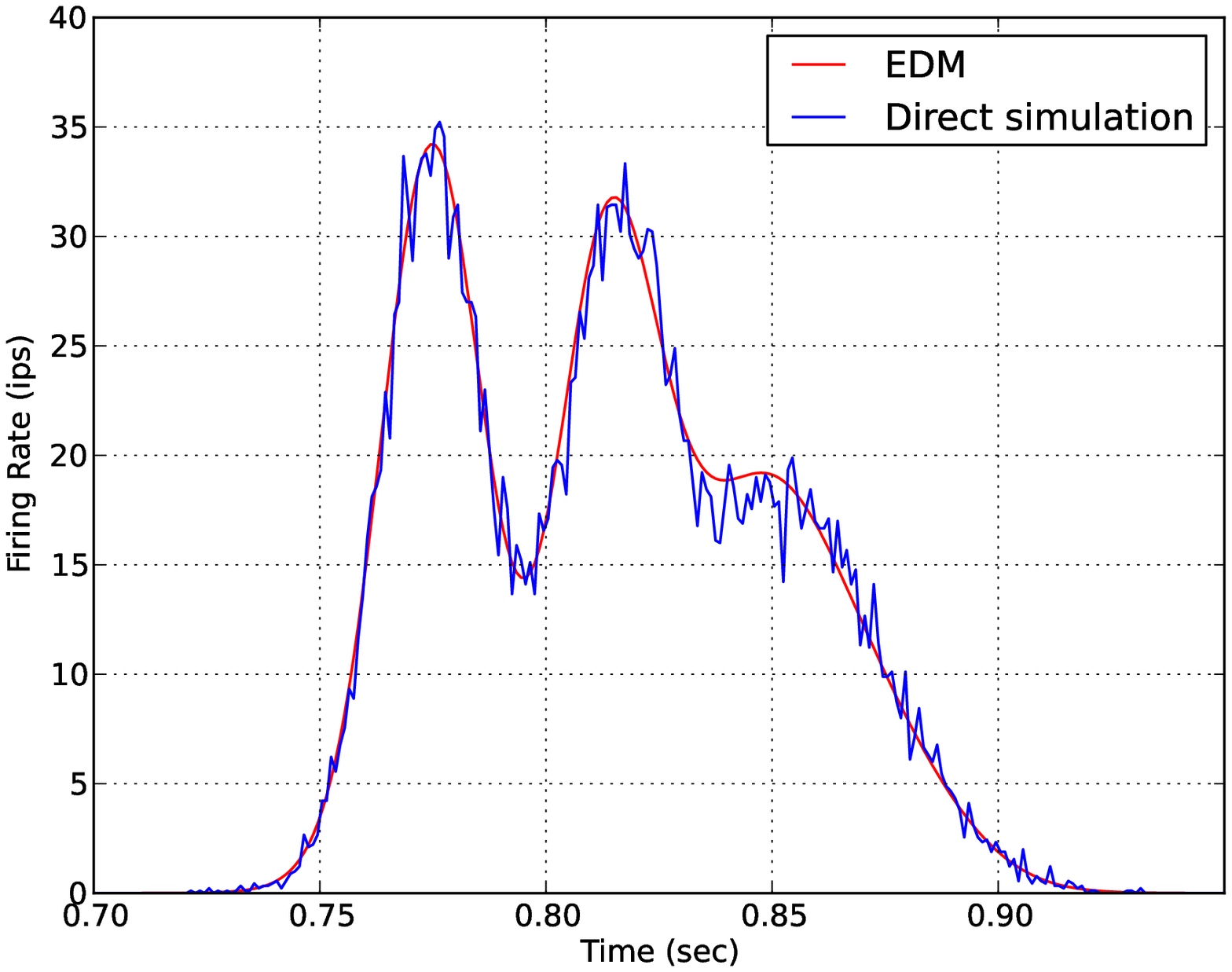}
\end{center}

\caption{Firing rate of a population of neurons, in response to a sinusoidal
input, calculated by direct simulation of a population of 9,000 IF neurons
(grey line) and by integrating a population equation (black line,
Equation~\ref{eq:J}). Neurons in this populations were
independent of each other (i.e., they had no feedback; G=0 in
Equation~\ref{eq:J}).}

\label{fig:firingRateSinusoidalRNFNI}
\end{figure}

The black line in Figure~\ref{fig:firingRateSinusoidalRNFNI} shows the
average firing rate per neuron, $r(t)$, calculated using the EDM,
Equation~\ref{eq:r}.  The grey line shows the average firing rate per neuron
calculated from a direct simulation of a population of 9,000 IF neurons, as
the proportion of cells with voltage at threshold. Note the almost perfect
match between these firing rates.

\begin{figure}
\begin{center}
\includegraphics[width=5in]{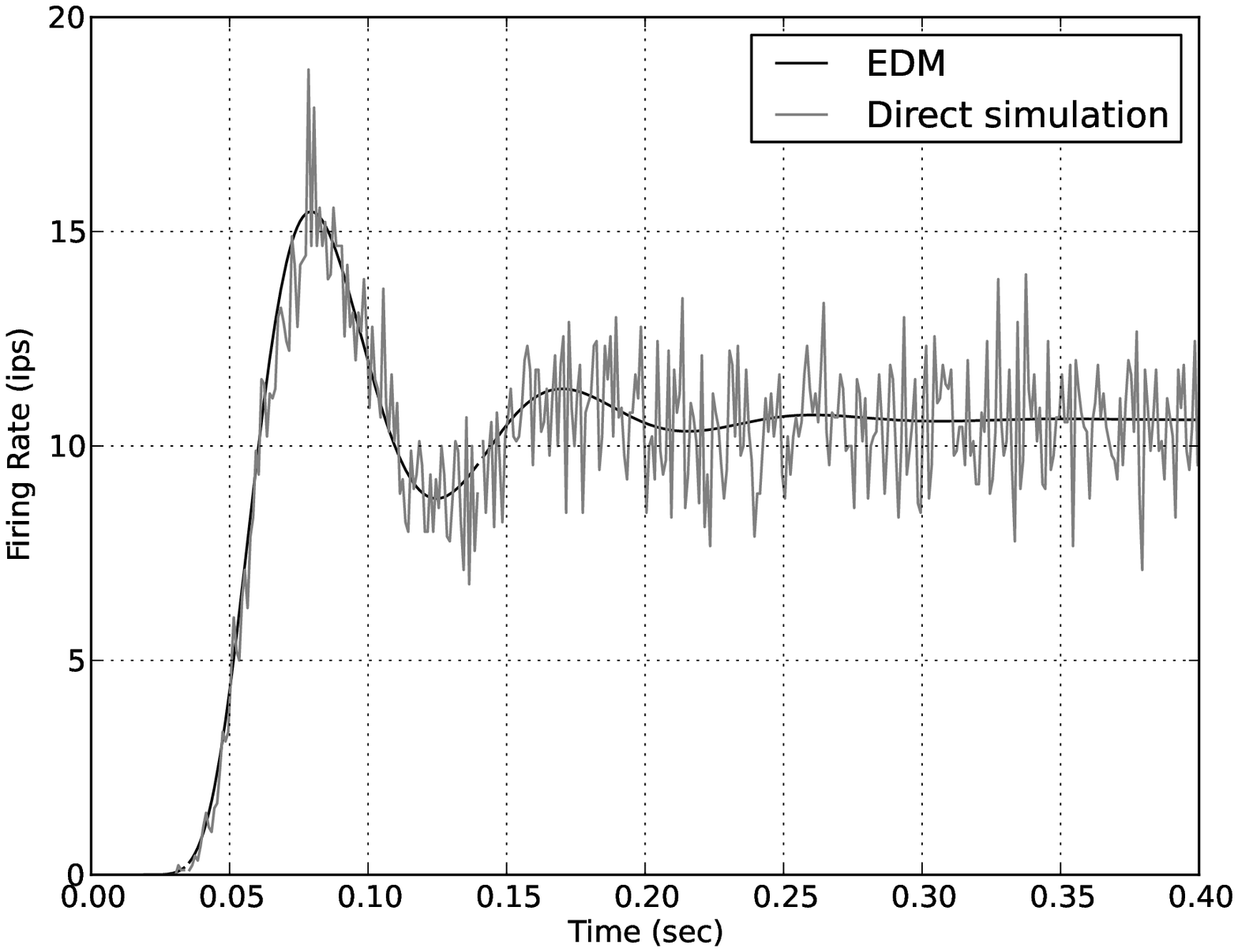}
\end{center}

\caption{Firing rate of a population of 9,000 IF neurons, in response to a
step input at time zero. Same format as in
Figure~\ref{fig:firingRateSinusoidalRNFNI}}

\label{fig:firingRateStepRNFNI}
\end{figure}

\begin{figure}
\begin{center}
\includegraphics[width=5in]{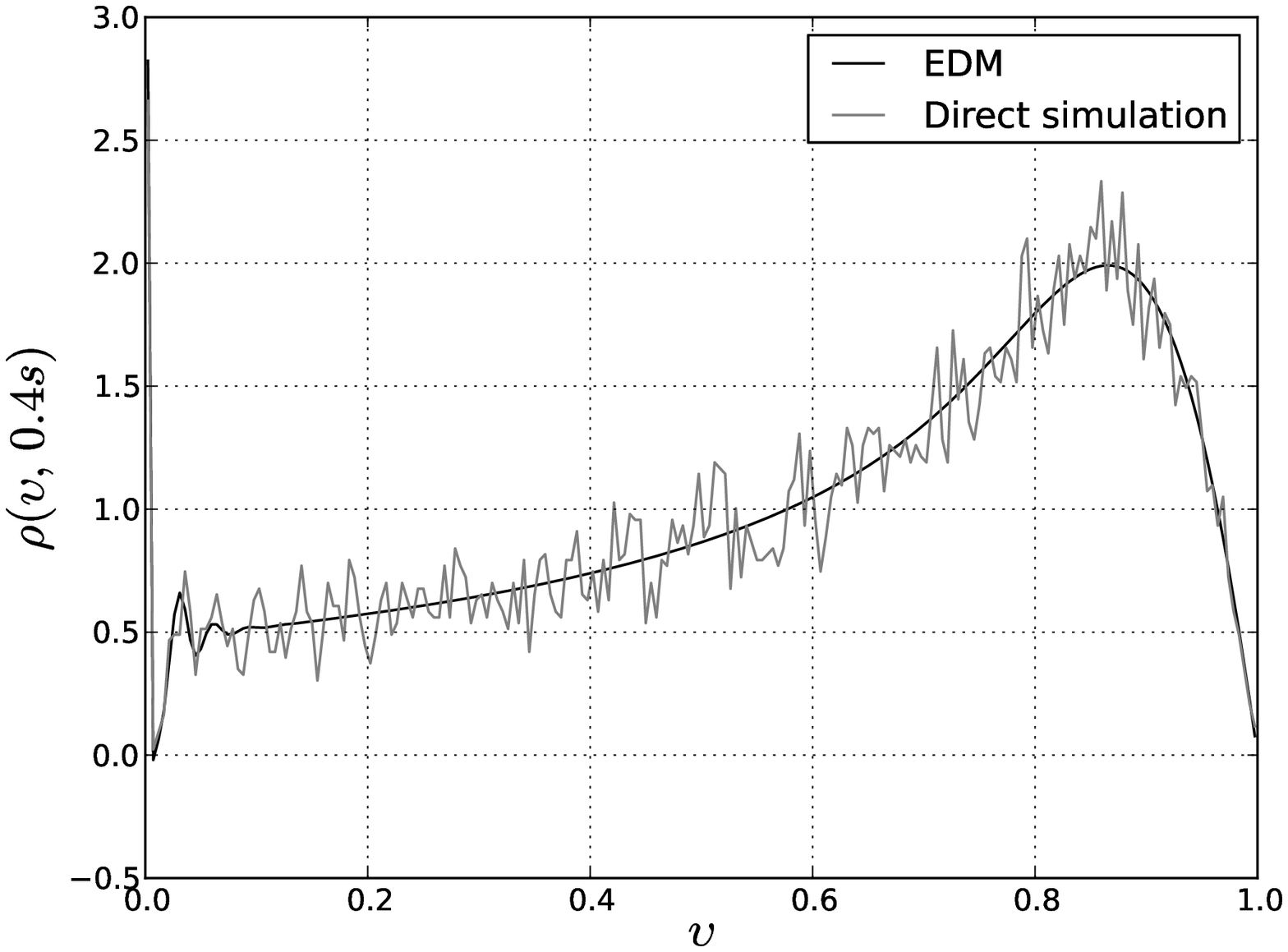}
\end{center}

\caption{Ensemble pdf, $\rho(\upsilon,t)$ at 0.4 seconds in the neurons of the
population of Figure~\ref{fig:firingRateStepRNFNI} in response of a step input
at time zero. This pdf was calculated by integrating the differential equation
of an EDM (Equation~\ref{eq:rhoDiffEq}, black line) and approximated using the
histogram of the voltages of the simulated neurons (grey line).}

\label{fig:rhosStepRNFNI}
\end{figure}

Figure~\ref{fig:firingRateStepRNFNI} is as
Figure~\ref{fig:firingRateSinusoidalRNFNI} but for a step input current that
jumps from 0 to 800 impulses per second at time $t=0$. Note that by 0.4
seconds after the step in the input current the average firing rate per
neuron has reached a new steady state around 10 impulses per second, as
revealed by the EDM (black line in Figure~\ref{fig:firingRateStepRNFNI}) and
by the direct simulation of a population of 9,000 IF neurons (grey line, in
Figure~\ref{fig:firingRateStepRNFNI})).  Figure~\ref{fig:rhosStepRNFNI} shows
the pdf of voltages at this new steady state calculated by the EDM,
Equation~\ref{eq:r} (black line in Figure~\ref{fig:rhosStepRNFNI}) and
approximated using the histogram of voltages at 0.4 ms of a direct simulation
of 9,000 IF neurons (grey line in Figure~\ref{fig:rhosStepRNFNI}).

\subsubsection{One Population with Feedback}

\begin{figure}
\begin{center}
\includegraphics[width=5in]{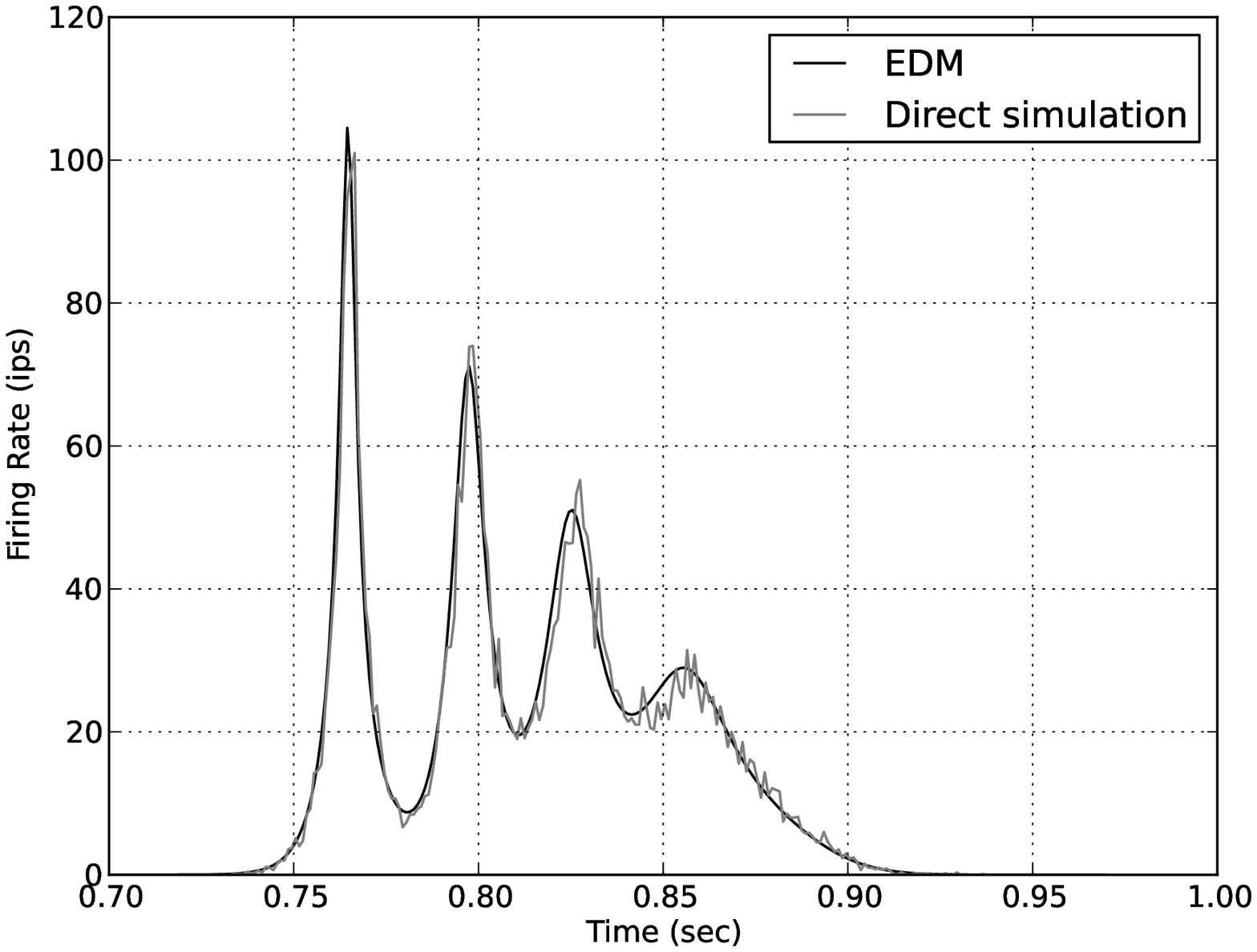}
\end{center}

\caption{Firing rate of a population of 9,000 neurons, in response to a
sinusoidal input, as in Figure~\ref{fig:firingRateSinusoidalRNFNI}, but each
neuron in this population is connected with ten presynaptic neurons, and all
of these neurons are excitatory (i.e., G=10 and f=1.0 in
Equation~\ref{eq:J}).}

\label{fig:firingRateSinusoidalRWFNI}
\end{figure}

The previous Figures showed results from the simulation of a single
population of independent neurons. Figure~\ref{fig:firingRateSinusoidalRWFNI}
is as Figure~\ref{fig:firingRateSinusoidalRNFNI}, but shows the average
firing rate per neuron in a population where each neuron receives excitatory
inputs from ten other neurons in the population (i.e., G=10 and f=1 in
Equation~\ref{eq:J}). For comparison, the dashed line shows the average
firing rate per neuron from the population without feedback.

\begin{figure}
\begin{center}
\includegraphics[width=5in]{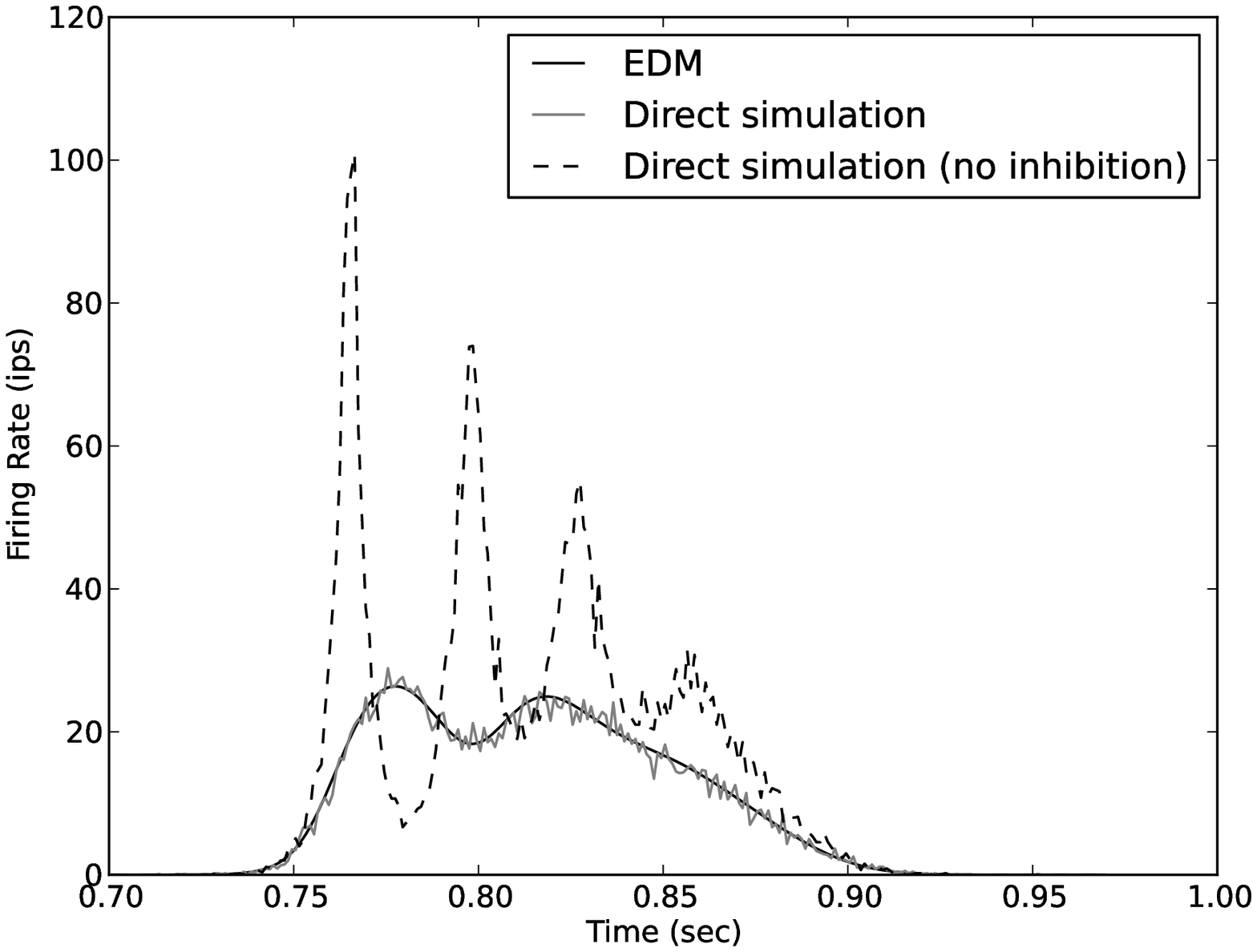}
\end{center}

\caption{Same as Figure~\ref{fig:firingRateSinusoidalRWFNI} but for a
population of neurons where 80\% of the presynaptic neurons to a given neuron
are inhibitory (i.e., G=10 and f=1-0.8 in Equation~\ref{eq:J}). For
comparison, the dashed line shows the average firing rate per neuron in the
population with feedback but not inhibition of
Figure~\ref{fig:rhosStepRNFNI}.}

\label{fig:firingRateSinusoidalRWFWI}
\end{figure}

Figure~\ref{fig:firingRateSinusoidalRWFWI} demonstrates the effect of
inhibitory feedback in the population of
Figure~\ref{fig:firingRateSinusoidalRWFNI} by changing the fraction of
inhibitory input neurons from 0\% to 80\% (by changing f=1-0 to f=1-0.8, and
still using G=10 in Equation~\ref{eq:J}).

\subsection{A Network of Populations}
\label{sec:networkOfPopulations}

\begin{figure}
\begin{center}
\includegraphics[width=5in]{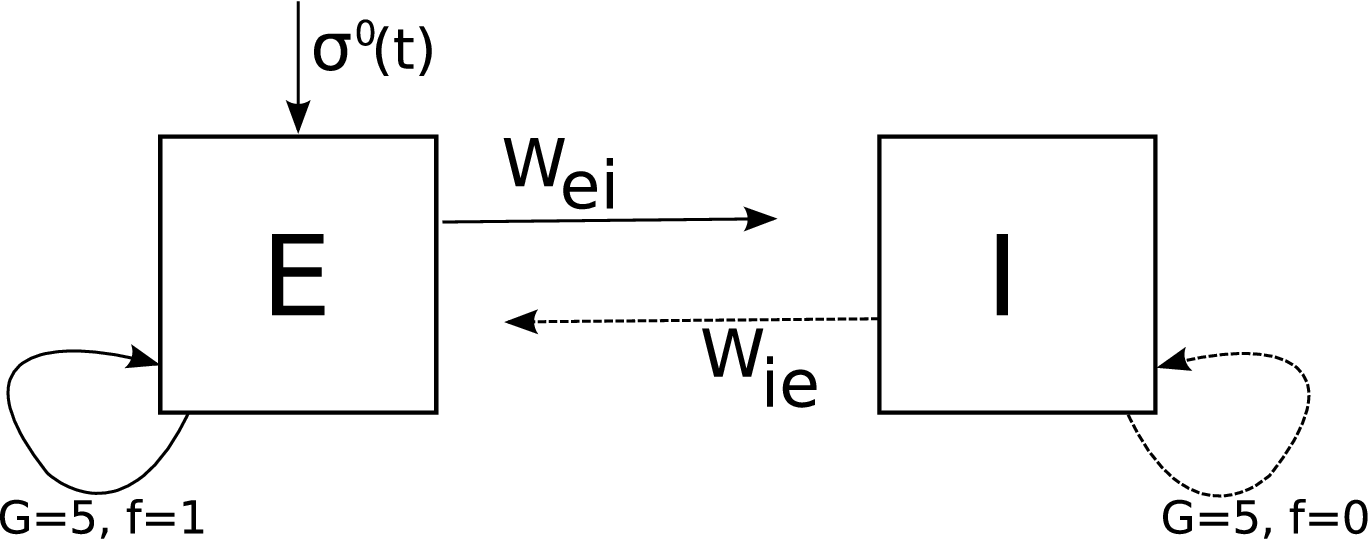}
\end{center}

\caption{Simulated network of two populations of IF neurons. A sinusoidal
input, $\sigma^0(t)$, was applied to the excitatory population. This
population contained excitatory feedback (each neuron in the population
received excitatory input from five other neurons in the population; G=5, f=1
in Equation~\ref{eq:J}). The average firing rate per neuron of the excitatory
population, scaled by the coefficient $W_{ei}=50$, was the excitatory input
for the inhibitory population. This population contained inhibitory feedback
(each neuron in the population received inhibitory input from five other
neurons in the population; G=5, f=0 in Equation~\ref{eq:J}). The average
firing rate per neuron of the inhibitory population, scaled by a coefficient
$W_{ie}=15$, was the inhibitory input for the excitatory population.
Excitatory/inhibitory connections are shown by solid/dashed lines.}

\label{fig:twoPopulations}
\end{figure}

The previous sections evaluated EDMs in a single population of neurons. Here
we evaluate EDMs of excitatory and inhibitory populations combined in the
network of populations shown in Figure~\ref{fig:twoPopulations}. The network
is driven by an excitatory input to the excitatory population. This
population has excitatory feedback (i.e., G=5 and f=1 in
Equation~\ref{eq:J}), and its average firing rate per neuron output,
scaled by a constant $W_{ei}=50$, drives the inhibitory population. The
inhibitory population has inhibitory feedback (i.e., G=5 and f=0 in
Equation~\ref{eq:J}),  and its average firing rate per neuron output
, scaled by a constant $W_{ie}=15$, modulates the activity of the excitatory
population.

\begin{figure}
\begin{center}
\includegraphics[width=4in]{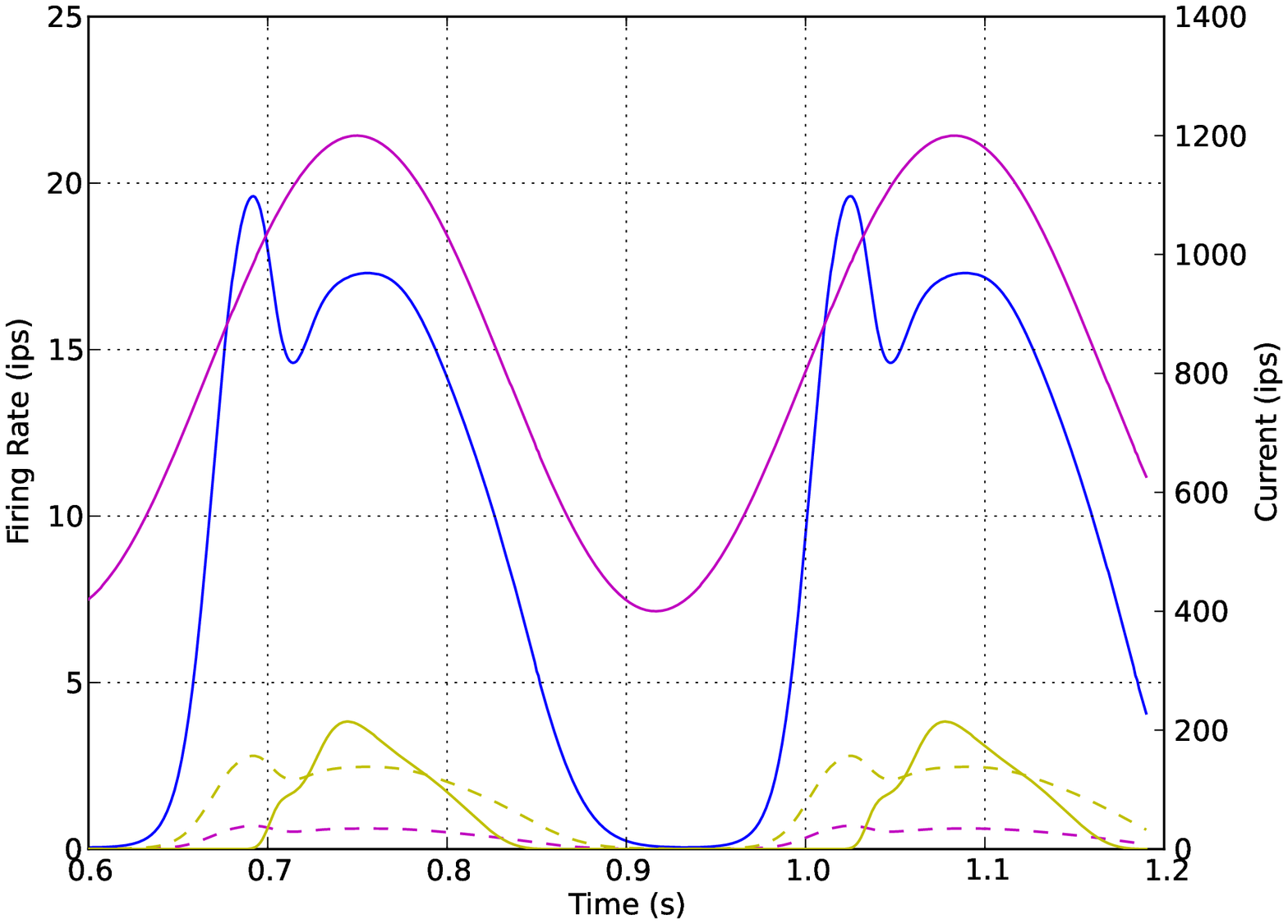}
\includegraphics[width=4in]{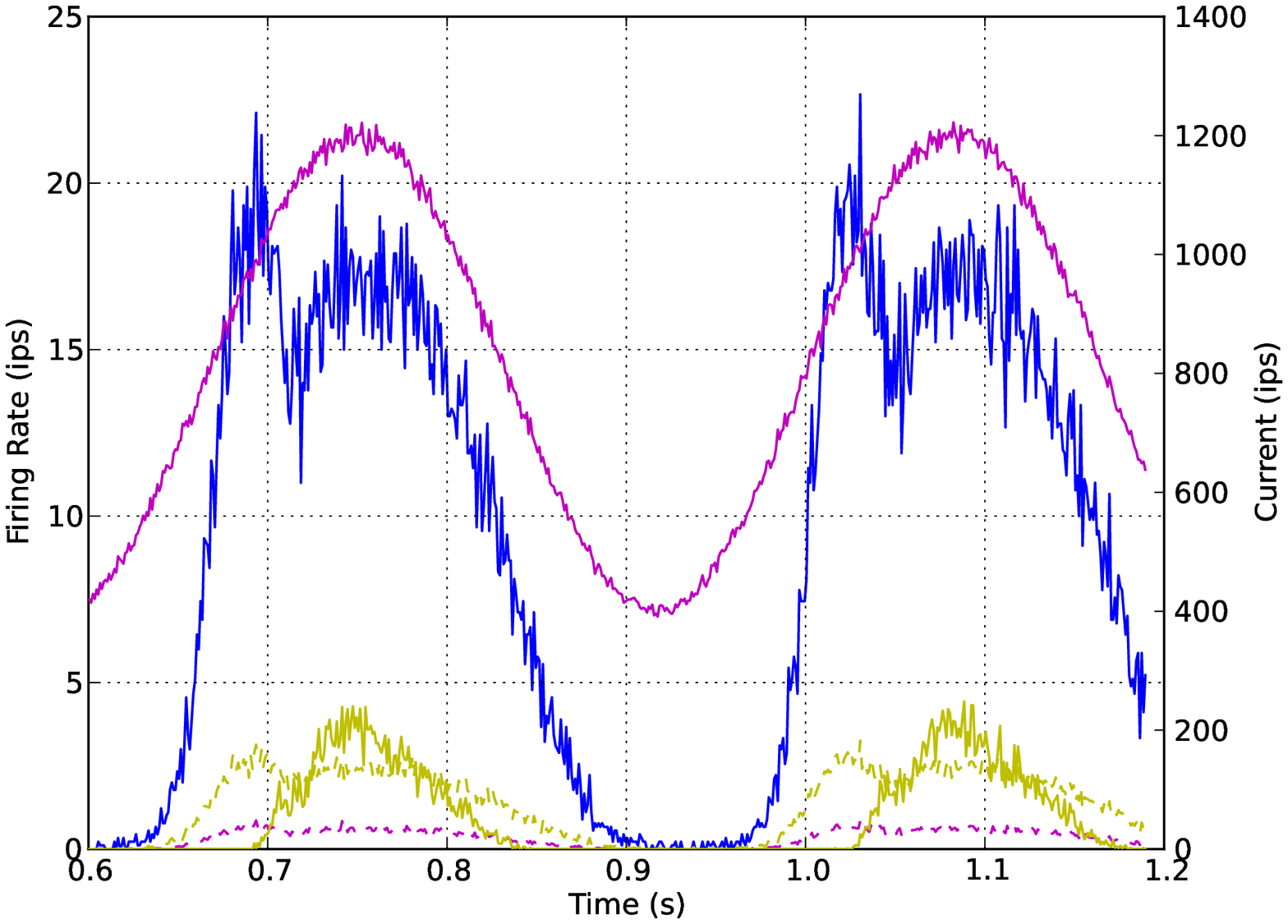}
\end{center}

\caption{The top and bottom panels show the activities of the excitatory
population represented by the population equation, and by the simulation of
9,000 IF neurons. The blue line, scaled on the left axis, plots the firing
rate of the population. The magenta and yellow lines represent excitatory and
inhibitory currents, respectively; and the solid and dashed lines represent
external and feedback currents, respectively. The currents are scaled on the
right axis. Note that the similarity between the average firing rate per
neuron obtained from
the population equation and from the direct simulation (blue lines in the top
and bottom panels).}

\label{fig:excitatoryPopulation}
\end{figure}

\begin{figure}
\begin{center}
\includegraphics[width=5in]{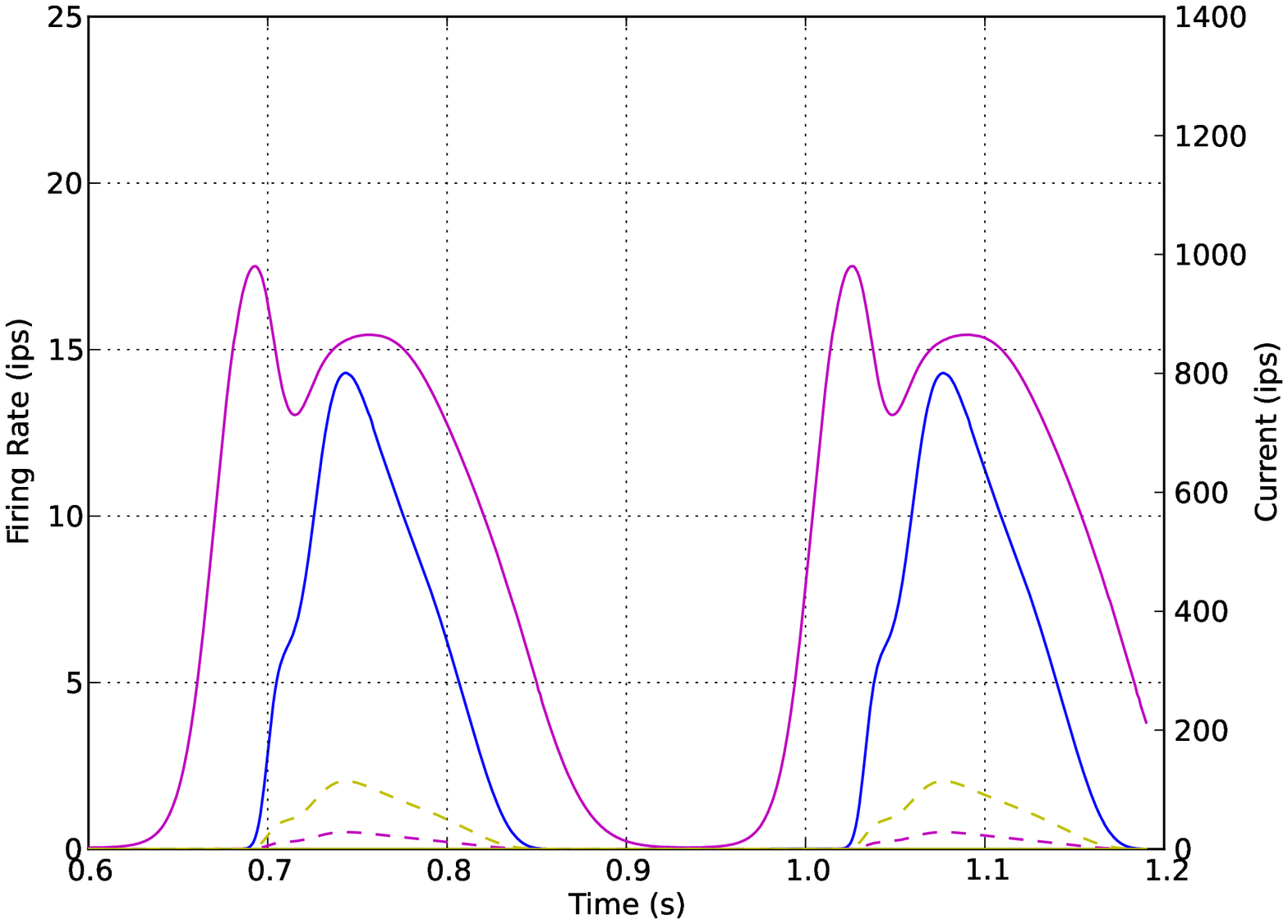}
\includegraphics[width=5in]{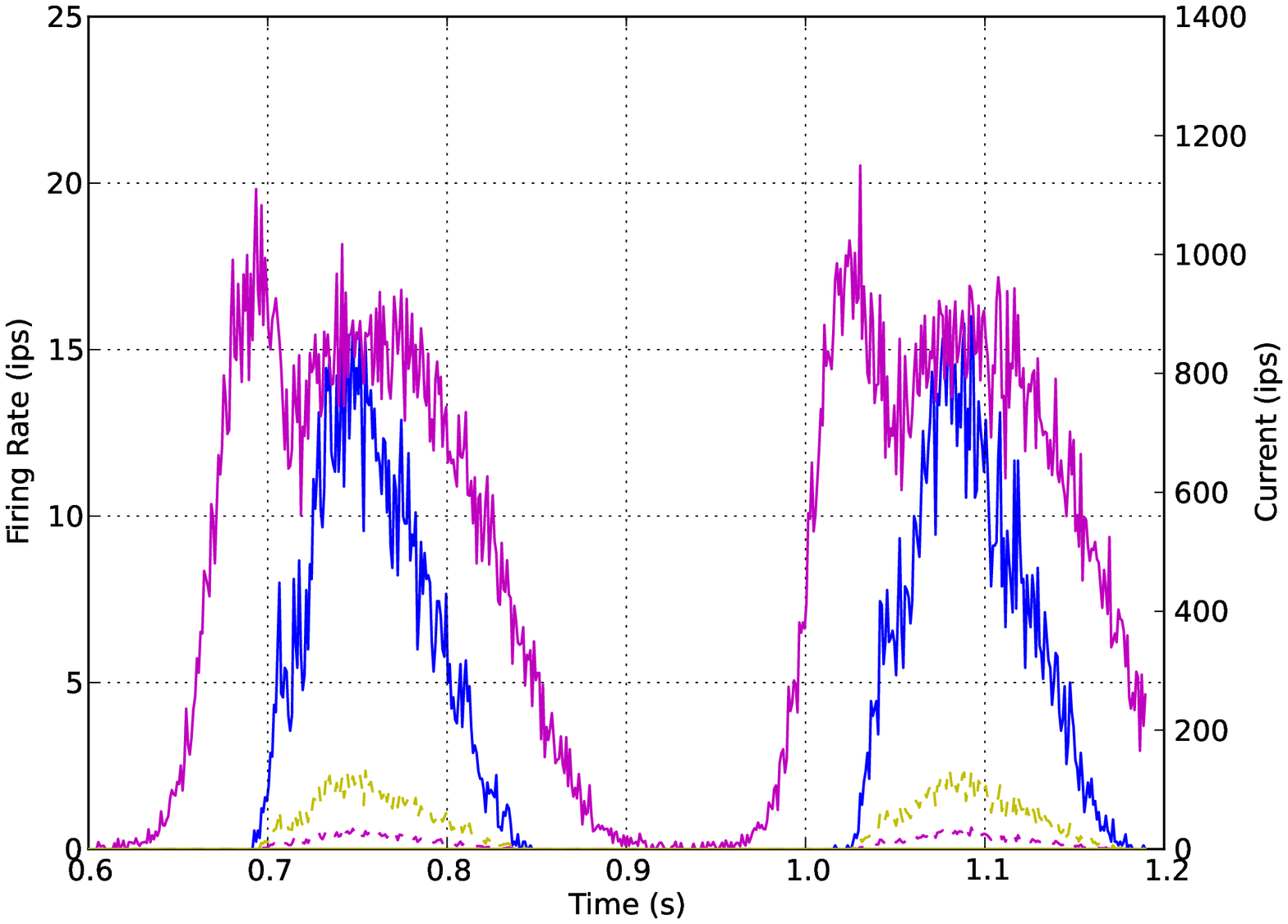}
\end{center}

\caption{Same as Figure~\ref{fig:excitatoryPopulation} but for the inhibitory
population in the two-populations model.}

\label{fig:inhibitoryPopulation}
\end{figure}

Figures~\ref{fig:excitatoryPopulation} and~\ref{fig:inhibitoryPopulation}
show the activity in the excitatory and inhibitory populations, respectively
in the network of Figure~\ref{fig:twoPopulations}. The upper panels in these
figures show the population activities computed by the EDMs and the bottom
panels the activity derived from the direct simulation of 9,000 IF neurons.
The blue curves, scaled along the left axis, show the average spike rate per
neuron in the population, the magenta and yellow curves, scaled along the
right axis, show the excitatory and inhibitory external inputs to the
population, respectively, and the magenta and yellow dashed curves, also scaled
along the right axis, show the excitatory and inhibitory feedback inputs to
the population. We should have performed the direct simulations with a larger
number of IF neurons to obtain smoother spike rates and currents.
Nevertheless, the activities computed by the EDMs are in close agreement with
those derived from the direct simulation.

\subsection{Partial Conclusions}

Given a dynamical model of a neuron, we now know how to derive an EDM for a
population of such neurons. For an IF model of a neuron, here we have shown
that EDMs accurately approximate population activity (i.e., the pdf of the
trans-membrane voltage, $\rho(\upsilon, t)$, and the average firing rate per
neuron, $r(t)$). The next step in this sub project is to estimate connectivity
parameters (e.g., $W_{ie}$ and $W_{ei}$ in Figure~\ref{fig:twoPopulations})
from simulated data.

\pagebreak

\section{Reducing dimensionality in EDMs}
\label{sec:reducingDimOfEDMs}

In the previous section we showed that Ensemble Density Models (EDMs) accurately
approximated the average firing rate per neuron and the probability density
function (pdf) of direct simulations of ensembles of integrate-and-fire (IF)
neurons.  In networks of EDMs, we want to estimate connectivity parameters and
state variables (i.e., the pdfs of the different ensembles) from recorded
ensemble firing rates. The state space of each previously reported EDM
contained 210 variables. To make the estimation of parameters and state
variables in networks of EDMs feasible/efficient it would be helpful to find
low-dimensional approximations of EDMs. Here we report the approximation power
of one such low-dimensional approximation method. This method was inspired by
the moving basis technique in~\citet{knight00}.

\subsection{Method to find low-dimensional approximations of EDMs}

The evolution of the ensemble pdf, $\rho(\upsilon, t)$, is given by:

\begin{eqnarray}
\dot\rho(\upsilon, t)=Q(s(t))\rho(\upsilon, t)
\label{eq:rhoEvolution}
\end{eqnarray}

\noindent where $Q(s(t))$ is a differential operator that depends on the
stimulus $s(t)$. The normalized voltage $\upsilon$ in
Equation~\ref{eq:rhoEvolution} ranges in the unit interval. To numerically
solve this equation, we discretize $\upsilon$, $\{\upsilon_i=i/N: 1\leq i\leq
N\}$ and $\rho$, $\{\rho_i(t)=\rho(\upsilon_i-\Delta/2, t): 1\leq i\leq N\}$,
giving a discretization of Equation~\ref{eq:rhoEvolution}:

\begin{eqnarray}
\dot{\bm{\rho}}(t)=\hat Q(s(t))\bm{\rho}(t)
\label{eq:discRhoEvolution}
\end{eqnarray}

\noindent where $\hat Q(s(t))\in\mathbb{R}^{N\times N}$ is the 
matrix representation of the differential operator $Q(s(t)$.
Equation~\ref{eq:discRhoEvolution} is a system of $N$ differential equations.
The objective of the dimensionality reduction method described below is to
approximate the evolution of $\bm{\rho}(t)$ using a system of $M$ differential
equations, where $M\ll N$. For this, the ensemble pdf $\bm\rho(t)$ is
represented in a basis of eigenvectors of the differential matrix $\hat
Q(s(t)$, where many coefficients in the new representation can be discarded
without much loss in approximation power.

\subsubsection{Representing the ensemble pdf in a new basis}

Let $\{\phi_n(s(t)): 0\leq n<N\}$ and $\{\lambda_n(s(t)): 0\leq n<N\}$ be the
eigenvectors and eigenvalues, respectively, of $\hat Q(s(t))$: 

\begin{eqnarray}
\hat Q(s(t))\bm\phi_n(s(t))=\lambda_n(s(t))\bm\phi_n(s(t))
\label{eq:eigenEq}
\end{eqnarray}

\noindent or in matrix notation:

\begin{eqnarray}
\hat Q(s(t))\Phi(s(t))=\Phi(s(t))\Lambda(s(t))
\label{eq:eigenEqMatrix}
\end{eqnarray}

\noindent where $\bm\phi_n(s(t))$ is the nth column of the matrix $\Phi(s(t))$
and $\lambda_n(s(t))$ is the nth diagonal element of the diagonal matrix
$\Lambda(s(t))$.

Assuming the eigenvectors are linearly independent, we represent
$\bm\rho(t)$ as:

\begin{eqnarray}
\bm\rho(t)=\sum_{n=1}^Na_n(t)\bm\phi_n(s(t))
\label{eq:rhoInEigenBasis}
\end{eqnarray}

\noindent or in matrix notation:

\begin{eqnarray}
\bm\rho(t)=\Phi(s(t))\,\mathbf{a}(t)
\label{eq:rhoInEigenBasisMatrix}
\end{eqnarray}

From Equations~\ref{eq:rhoInEigenBasisMatrix} and~\ref{eq:eigenEqMatrix} it
follows:

\begin{eqnarray}
\hat Q(s(t))\bm\rho(t)=\hat Q(s(t))\Phi(s(t))\,\mathbf{a}(t)=\Phi(s(t))\Lambda(s(t))\,\mathbf{a}(t)
\label{eq:rhoDotInEigenBasisMatrix}
\end{eqnarray}

Using a backward difference to
approximate the time derivative of $\bm\rho(t)$:

\begin{eqnarray}
\dot{\bm\rho}(t)=\frac{\bm\rho(t)-\bm\rho(t-\Delta t)}{\Delta t}
\label{eq:discretizationRhoDot}
\end{eqnarray}

\noindent in Equation~\ref{eq:discRhoEvolution} we obtain:

\begin{eqnarray}
\bm\rho(t)-\Delta t\,\hat Q(s(t))\bm\rho(t)=\bm\rho(t-\Delta t)
\label{eq:derivationMotionA1}
\end{eqnarray}

Now applying Equations~\ref{eq:rhoInEigenBasisMatrix}
and~\ref{eq:rhoDotInEigenBasisMatrix} to Equation~\ref{eq:derivationMotionA1}
we get:

\begin{eqnarray}
\left[\Phi(s(t))\left(I-\Delta t \Lambda(s(t))\right)\right]\mathbf{a}(t)=\Phi(s(t-\Delta t)\mathbf{a}(t-\Delta t)
\label{eq:derivationMotionA2}
\end{eqnarray}

\noindent or:

\begin{eqnarray}
\mathbf{a}(t)=\left[\left(I-\Delta t \Lambda(s(t))\right)^{-1}\Phi(s(t))^{-1}\Phi(s(t-\Delta t)\right]\mathbf{a}(t-\Delta t)
\label{eq:motionA}
\end{eqnarray}

Equation~\ref{eq:motionA} provides the evolution of the coefficients
$\mathbf{a}(t)$ in Equation~\ref{eq:rhoInEigenBasisMatrix}. It just expresses
the evolution of the ensemble pdf in Equation~\ref{eq:discRhoEvolution} in
another basis.  It is an exact formula (i.e., it is not an approximation) and
has the same dimensionality as Equation~\ref{eq:discRhoEvolution}. The
importance of Equation~\ref{eq:motionA} is that one can approximate the
evolution of the ensemble pdf by discarding many components of $\mathbf{a}(t)$,
as we explain next.

\subsubsection{Reducing dimensionality in the new basis}

To understand why one can discard many coefficients in the representation of
the ensemble pdf of Equation~\ref{eq:rhoInEigenBasisMatrix} without much loss
in approximation power, consider the case where the stimulus, $s(t)$, is
constant. In such a case, the eigenvectors and eigenvalues will
neither depend on the stimulus; i.e., $\Phi(s(t))=\Phi$ and
$\Lambda(s(t))=\Lambda$. Then
Equations~\ref{eq:discRhoEvolution}
and~\ref{eq:rhoInEigenBasisMatrix} reduce to:

\begin{eqnarray}
\dot{\bm{\rho}}(t)=\hat Q\bm{\rho}(t)
\label{eq:discRhoEvolutionConstantStim}
\end{eqnarray}

\noindent and:

\begin{eqnarray}
\bm\rho(t)=\Phi\,\mathbf{a}(t)
\label{eq:rhoInEigenBasisMatrixConstantStim}
\end{eqnarray}

Taking derivatives in Equation~\ref{eq:rhoInEigenBasisMatrixConstantStim} we
obtain:

\begin{eqnarray}
\dot{\bm\rho}(t)=\Phi\,\dot{\mathbf{a}}(t)
\label{eq:derRhoInEigenBasisMatrixConstantStimV1}
\end{eqnarray}

\noindent and substituing Equation~\ref{eq:rhoInEigenBasisMatrixConstantStim}
in Equation~\ref{eq:discRhoEvolutionConstantStim}, and applying
Equation~\ref{eq:eigenEqMatrix}, we get:

\begin{eqnarray}
\dot{\bm{\rho}}(t)=\hat
Q\Phi\,\mathbf{a}(t)=\Phi\Lambda\,\mathbf{a}(t)
\label{eq:derRhoInEigenBasisMatrixConstantStimV2}
\end{eqnarray}

Equating the right hand sides in
Equations~\ref{eq:derRhoInEigenBasisMatrixConstantStimV1}
and~\ref{eq:derRhoInEigenBasisMatrixConstantStimV2}, and pre multiplying by
$\Phi^{-1}$, gives:

\begin{eqnarray}
\dot{\mathbf{a}}(t)=\Lambda\,\mathbf{a}(t)
\label{eq:motionAConstantStim}
\end{eqnarray}

\noindent or:

\begin{eqnarray}
\dot{a_i}(t)=\lambda_i\,a_i(t)
\label{eq:motionOneAConstantStim}
\end{eqnarray}

\noindent with solution:

\begin{eqnarray}
a_i(t)=\exp(\lambda_i t)\;a_i(0)
\label{eq:solutionOneAConstantStim}
\end{eqnarray}

Thus, the absolute value of a low dimensional coefficient $a_i$ evolves as:

\begin{eqnarray}
|a_i(t)|=\exp(\Re(\lambda_i) t)\;|a_i(0)|
\label{eq:solutionAbsOneAConstantStim}
\end{eqnarray}

Because all eigenvalues have negative real part (with the exception of the zero
eigenvalue), the coefficients associated with non-zero eigenvalues will decay
to zero, and the speed of this decay will be proportional to the absolute value
of the real part of the corresponding eigenvalue.  Therefore, to achieve
dimensionality reduction in Equation~\ref{eq:motionA} we may discard those
coefficients associated with eigenvalues with larger absolute value of their
real part, since these coefficients will rapidly decay to zero.

\subsection{Evaluation of the method}

We first study how well low-dimensional EDMs approximate the average firing
rate per neuron in
direct stimulations (Section~\ref{sec:firingRates}) and then how they
approximate the ensemble pdf (Section~\ref{sec:rhos}). 

These studies were peformed in data simulated from the network of EDMs
illustrated in Figure~\ref{fig:twoPopulations}. The network was driven by an
excitatory sinusoidal input to the excitatory population (shown by the dotted
curve and scaled along the right axis in the top panel of
Figure~\ref{fig:rsAndKLsEPop}). The average firing per neuron in this
population, scaled by a constant $W_{ei}=50$, drove the inhibitory population
(shown by the dotted curve and scaled along the right axis in the top panel of
Figure~\ref{fig:rsAndKLsIPop}). In turn, the average firing rate per neuron in
the inhibitory population, scaled by a constant $W_{ie}=15$, inhibited the
excitatory population.  Both populations contained feedback (i.e., each neuron
received 10 inputs from ten other cells in the same population and 80\% of
these inputs were inhibitory).

\subsubsection{Firing Rates}
\label{sec:firingRates}

\begin{figure}
\begin{center}
\includegraphics[width=5in]{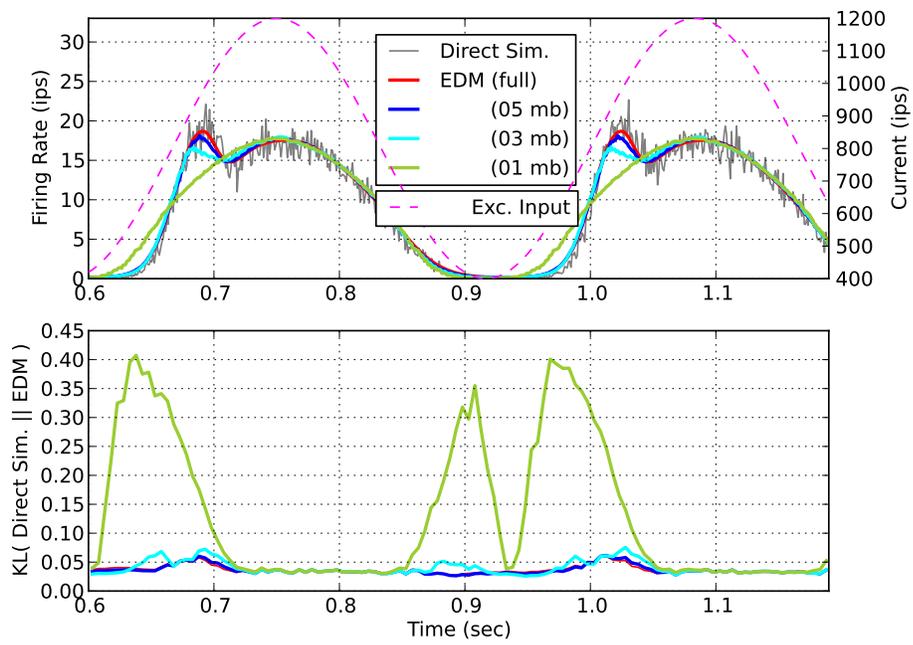}
\end{center}

\caption{Average firing rate per neuron (top) and KL divergence between the ensemble
pdf calculated by direct simulation and that caculated by EDMs (bottom) for the
excitatory ensemble.}

\label{fig:rsAndKLsEPop}
\end{figure}

\begin{figure}
\begin{center}
\includegraphics[width=5in]{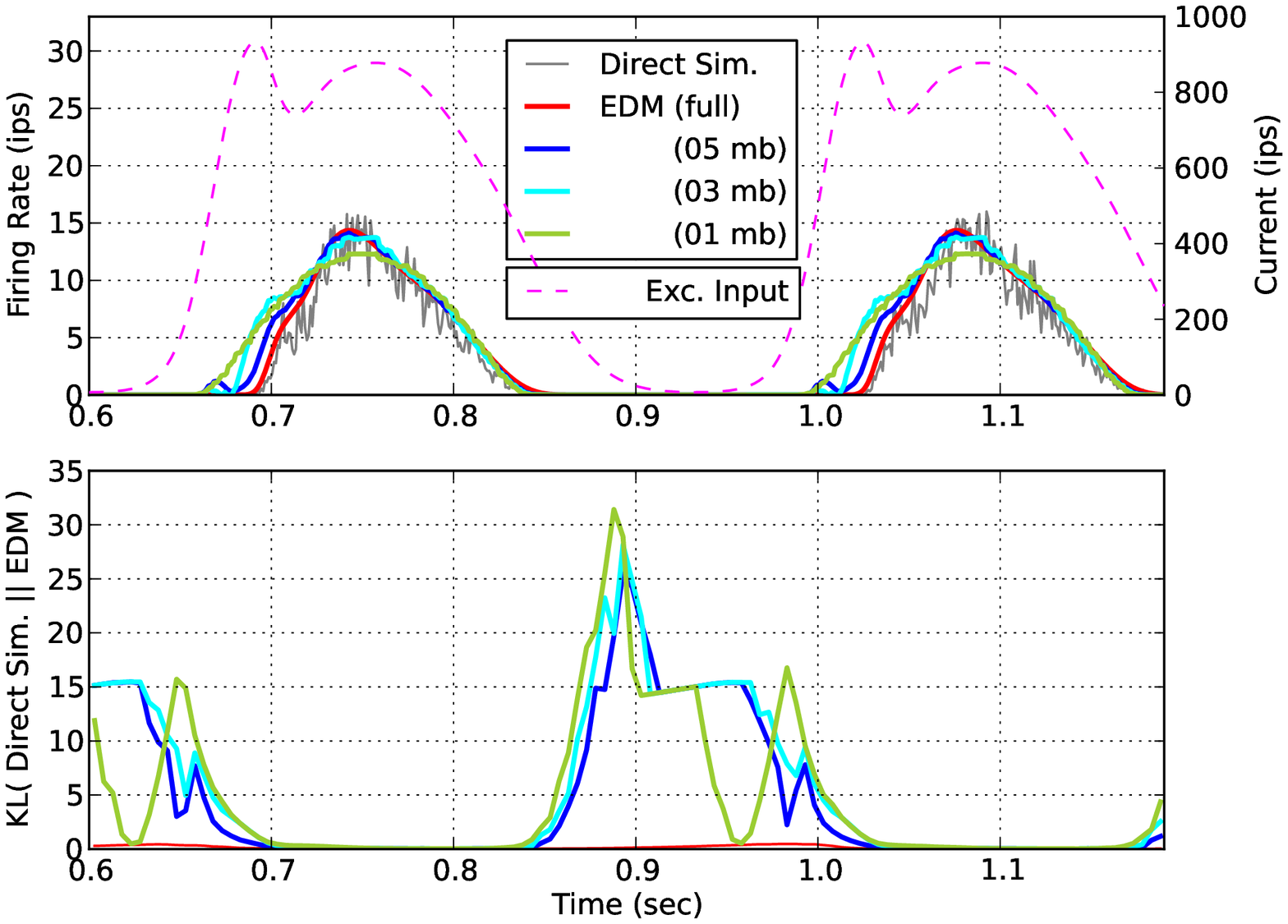}
\end{center}

\caption{Average firing rate per neuron (top) and KL divergence between the ensemble
pdf calculated by direct simulation and that caculated by EDMs (bottom) for the
inhibitory ensemble.}

\label{fig:rsAndKLsIPop}
\end{figure}

The top panels of Figures~\ref{fig:rsAndKLsEPop} and~\ref{fig:rsAndKLsIPop}
show the average firing rate per neuron obtained from direct simulation (grey
curve), from a full-dimensional EDM (red curve), and from low-dimensional EDMs
(the blue, cyan, and red curves correspond to 17, 5, and 1 moving basis,
respectively). The full dimensional EDM and its low dimensinal approximations
with 17 and 5 moving basis almost perfectly approximate the average firing rate
per neuron of the
direct simulation. The approximation power of the EDM with only one moving
basis is not as good, but it look reasonable.

\subsubsection{Ensemble Proability Density Functions}
\label{sec:rhos}

We compared the normalized histogram of number of directly simulated neuron per
voltage bin (i.e., the ensemble pdf from direct simulation) with the pdfs
calculated with EDMs (i.e., Equation~\ref{eq:discRhoEvolution}). For this we
computed at every time step the Kullback-Leibler (KL) divergence (in bits)
between the pdfs obtained by direct stimulation and those obtained from EDMs.
These KL divergences are shown in the bottom panels of
Figures~\ref{fig:rsAndKLsEPop} and~\ref{fig:rsAndKLsIPop}. 

We see that the pdfs obtained from EDMs were good approximation of the pdfs
from the direct simulation at times of large average firing rate per neuron (between 0.73 and
0.85 seconds and between 1.03 and 1.2 seconds in the bottom panel of
Figure~\ref{fig:rsAndKLsEPop}, and between 0.72 and 0.82 seconds and between
1.03 and 1.18 seconds in the bottom panel of Figure~\ref{fig:rsAndKLsIPop}).

At times of low averaged firing rate per neuron the difference between the pdfs
obtained from direct simulation and those obtained from EDMs were an order of
magnitude larger for the inhibitory than for the excitatory ensemble. This is
probably because the excitatory ensemble was driven by a large and smooth
sinusoidal input while the inhibitory ensemble was driven by the weaker and
non-smooth average firing rate per neuron of the excitatory ensemble. Also, at
most time points, the larger the number of moving basis in low-dimensional
EDMs, the better the EDM pdf approximated the pdf obtained from direct
simulation, as can be seen more clearly in the bottom panel of
Figure~\ref{fig:rsAndKLsIPop}.

\begin{figure}
\begin{center}
\includegraphics[width=5in]{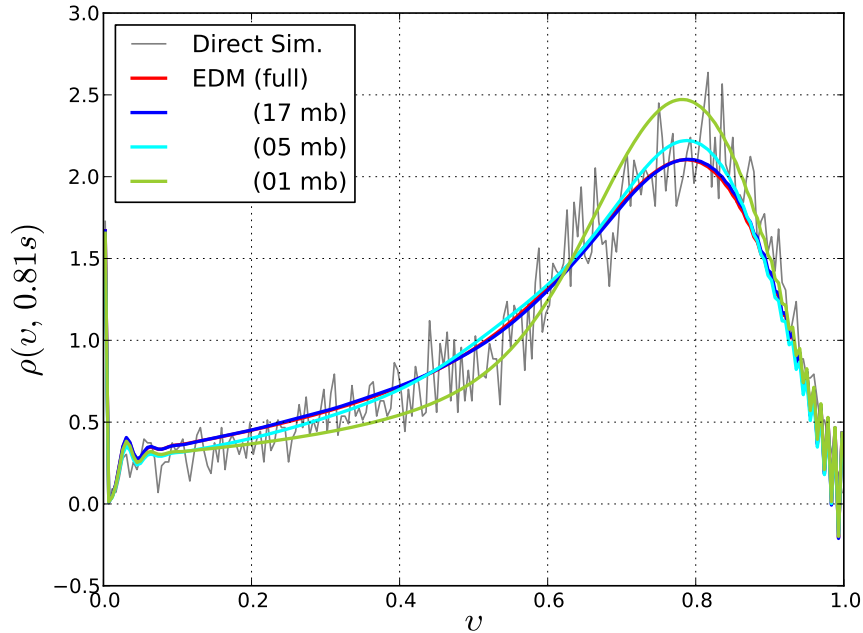}
\end{center}

\caption{Ensemble pdfs for the inhibitory ensemble at 0.81s.}

\label{fig:iRhosAt0.81}
\end{figure}

\begin{figure}
\begin{center}
\includegraphics[width=5in]{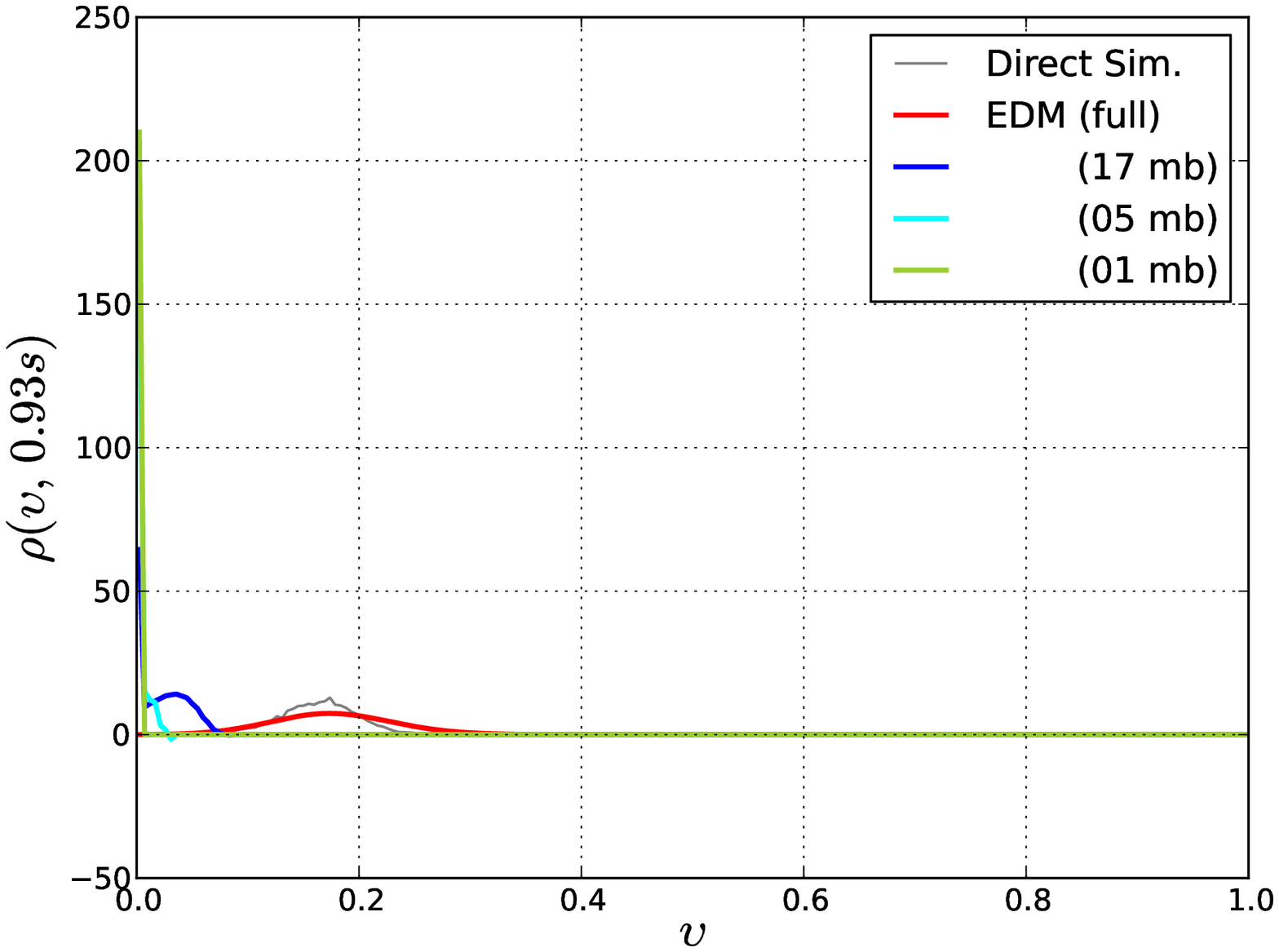}
\end{center}

\caption{Ensemble pdfs for the inhibitory ensemble 0.93s.}

\label{fig:iRhosAt0.93}
\end{figure}

\begin{figure}
\begin{center}
\includegraphics[width=5in]{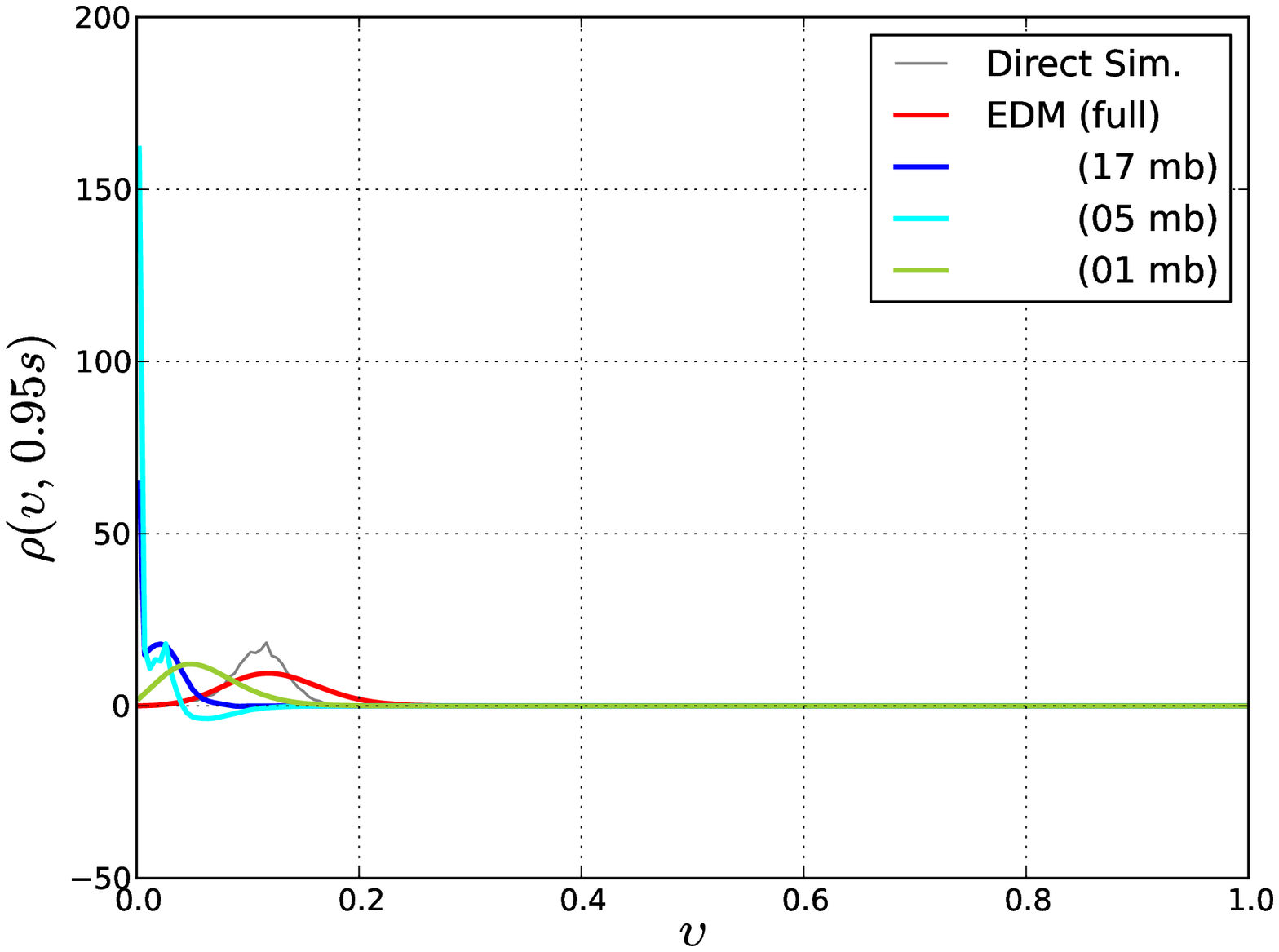}
\end{center}

\caption{Ensemble pdfs for the inhibitory ensemble 0.95s.}

\label{fig:iRhosAt0.95}
\end{figure}

\begin{figure}
\begin{center}
\includegraphics[width=5in]{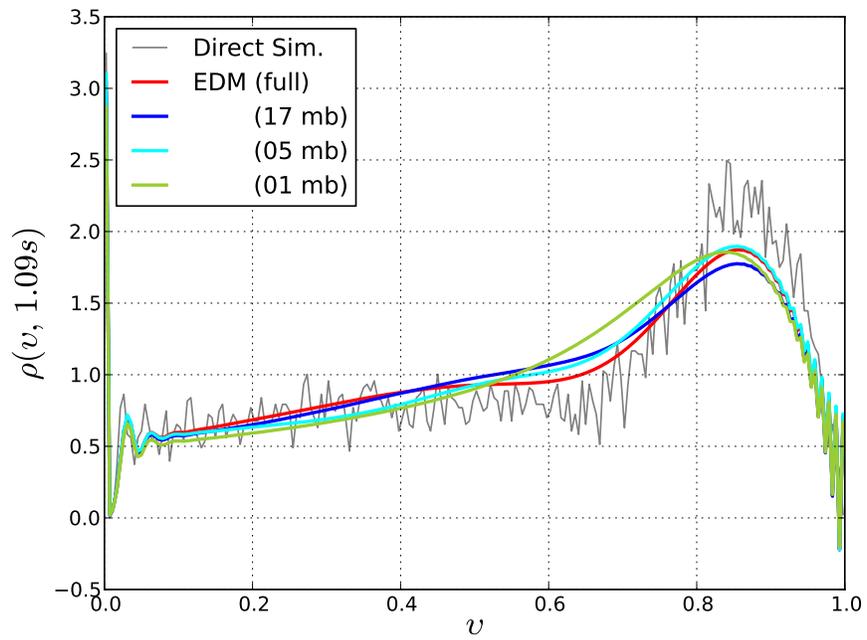}
\end{center}

\caption{Ensemble pdfs for the inhibitory ensemble 1.09s.}

\label{fig:iRhosAt1.09}
\end{figure}

To try to understand why the pdfs obtained from direct simulation were
different from those obtained from low-dimensional EDMs, we plotted these pdfs
for the inhibitory ensemble in an interval of low average firing rate per
neuron (from 0.81 seconds in Figure~\ref{fig:iRhosAt0.81} to 1.09 seconds in
Figure~\ref{fig:iRhosAt1.09}).  In the transition between weak to zero average
firing rate per neuron (from 0.81 seconds in Figure~\ref{fig:iRhosAt0.81} to
0.93 seconds in Figure~\ref{fig:iRhosAt0.93}) the low-dimensional pdfs moved
faster towards lowers voltage than the pdfs from the full-dimensional EDM and
those from direct simulation. Similarly, in the transition between zero to weak
average firing rate per neuron (from 0.95 seconds in
Figure~\ref{fig:iRhosAt0.95} to 1.09 seconds in Figure~\ref{fig:iRhosAt1.09})
the low-dimensional pdfs moved faster towards higher voltages than the pdfs from
the full-dimensional EDM and those from direct simulation. This suggests that
the moving basis discarded in the low-dimensional approximations of EDMs help
prevent the EDM pdf to transition too fast to and away from the pdf
corresponding to large average firing rate per neuron.

\begin{figure}
\begin{center}
\includegraphics[width=5in]{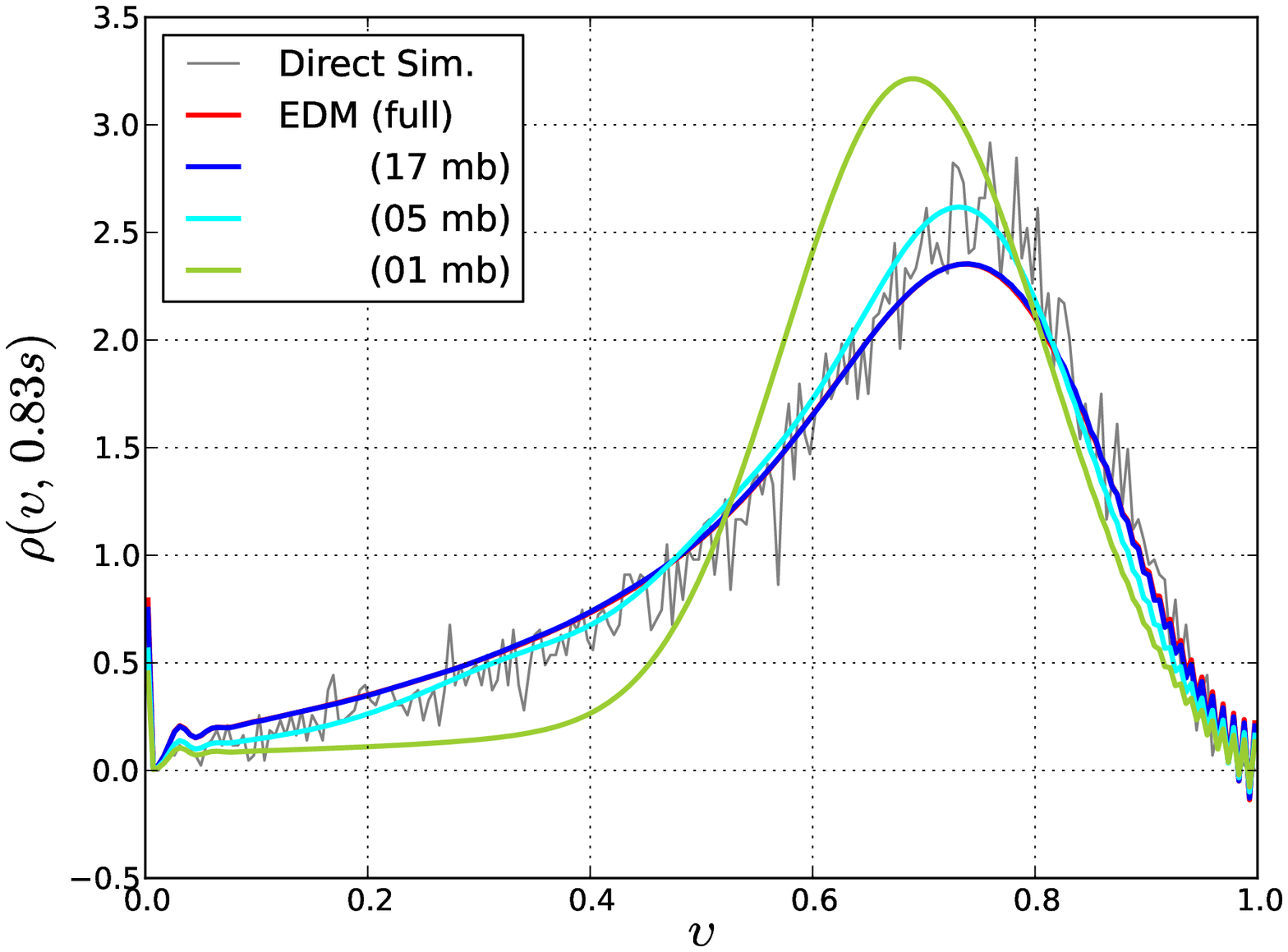}
\end{center}

\caption{Ensemble pdfs for the inhibitory ensemble at 0.83s.}

\label{fig:iRhosAt0.83}
\end{figure}

\begin{figure}
\begin{center}
\includegraphics[width=5in]{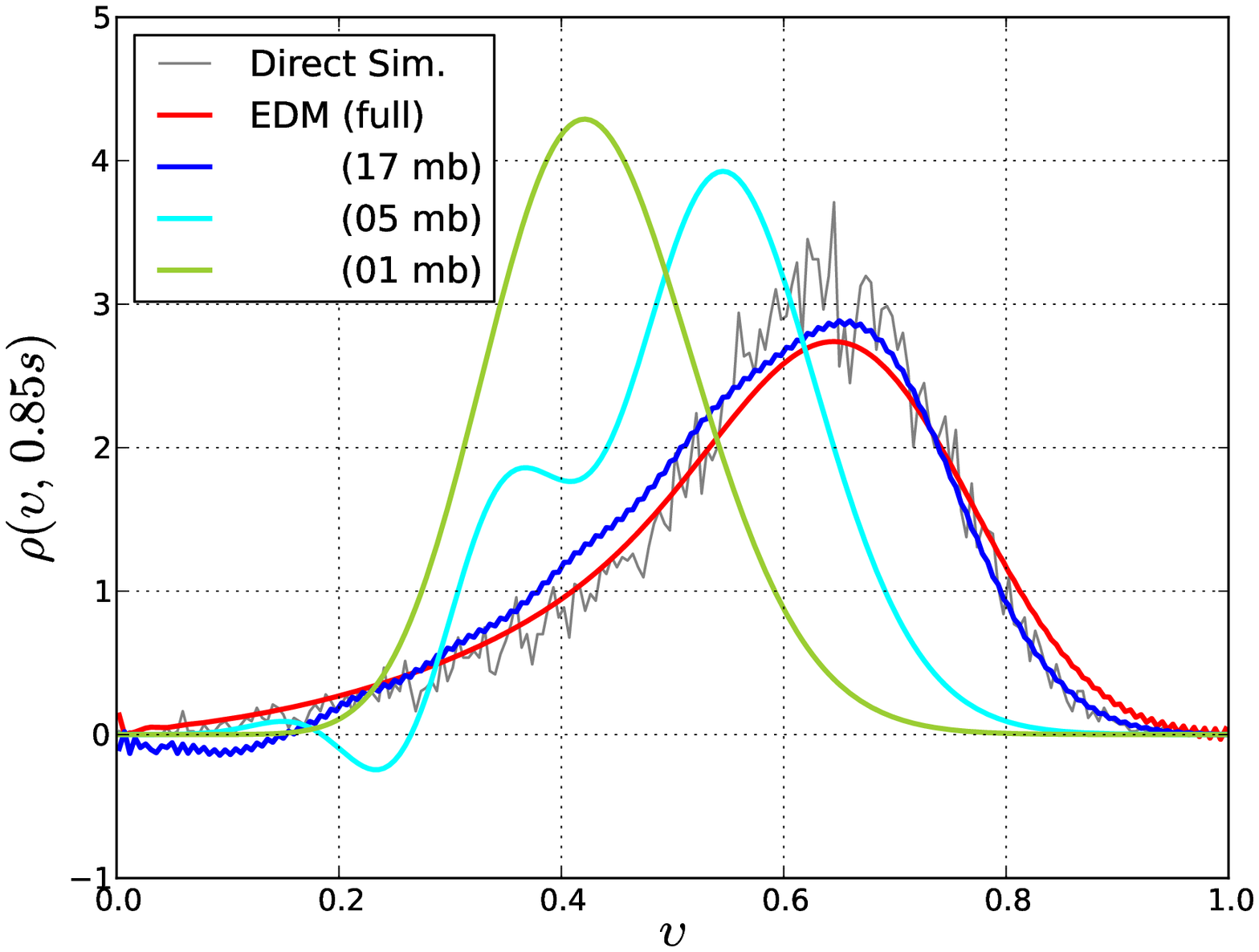}
\end{center}

\caption{Ensemble pdfs for the inhibitory ensemble 0.85s.}

\label{fig:iRhosAt0.85}
\end{figure}

\begin{figure}
\begin{center}
\includegraphics[width=5in]{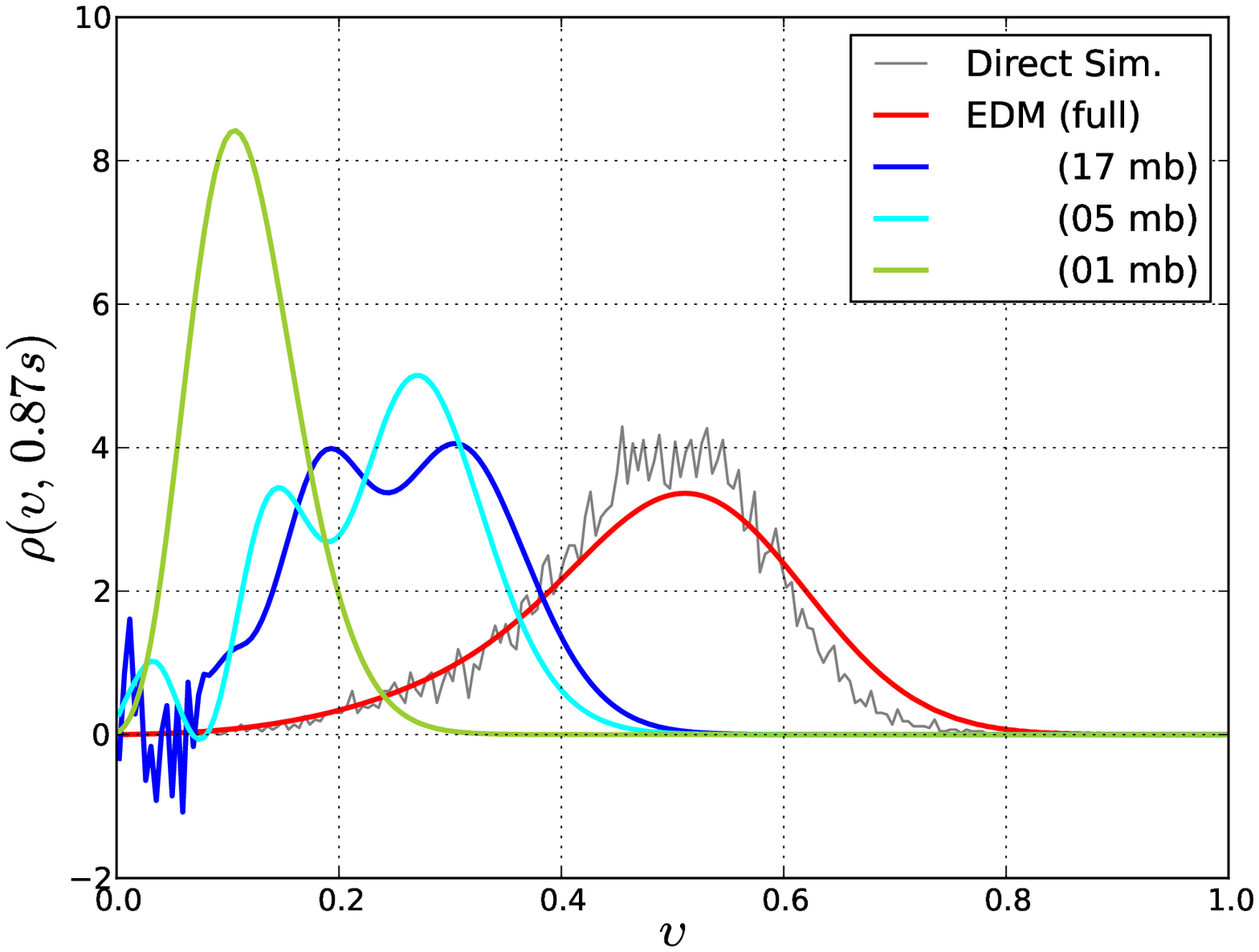}
\end{center}

\caption{Ensemble pdfs for the inhibitory ensemble 0.87s.}

\label{fig:iRhosAt0.87}
\end{figure}

\begin{figure}
\begin{center}
\includegraphics[width=5in]{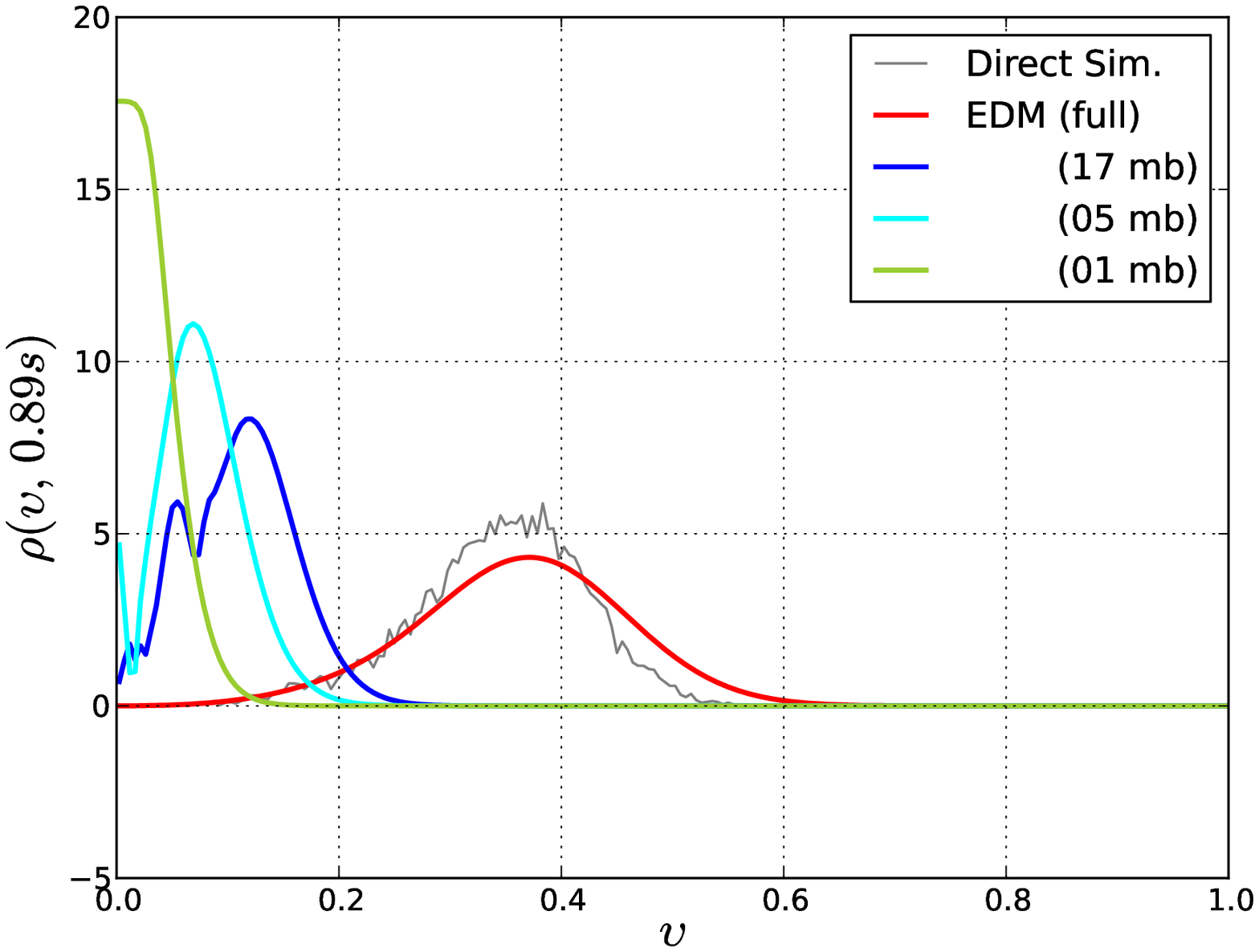}
\end{center}

\caption{Ensemble pdfs for the inhibitory ensemble 0.89s.}

\label{fig:iRhosAt0.89}
\end{figure}

\begin{figure}
\begin{center}
\includegraphics[width=5in]{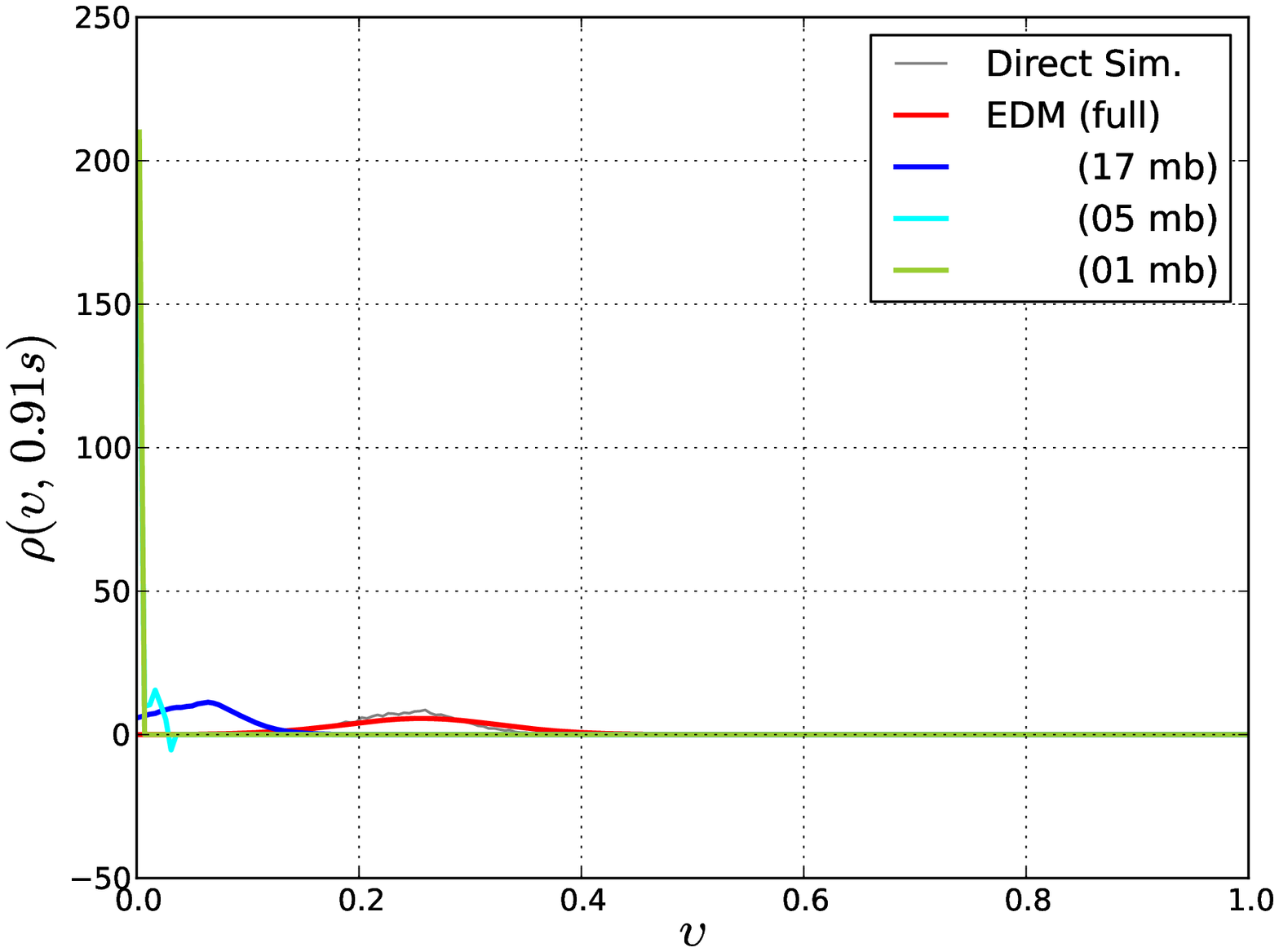}
\end{center}

\caption{Ensemble pdfs for the inhibitory ensemble 0.91s.}

\label{fig:iRhosAt0.91}
\end{figure}

\begin{figure}
\begin{center}
\includegraphics[width=5in]{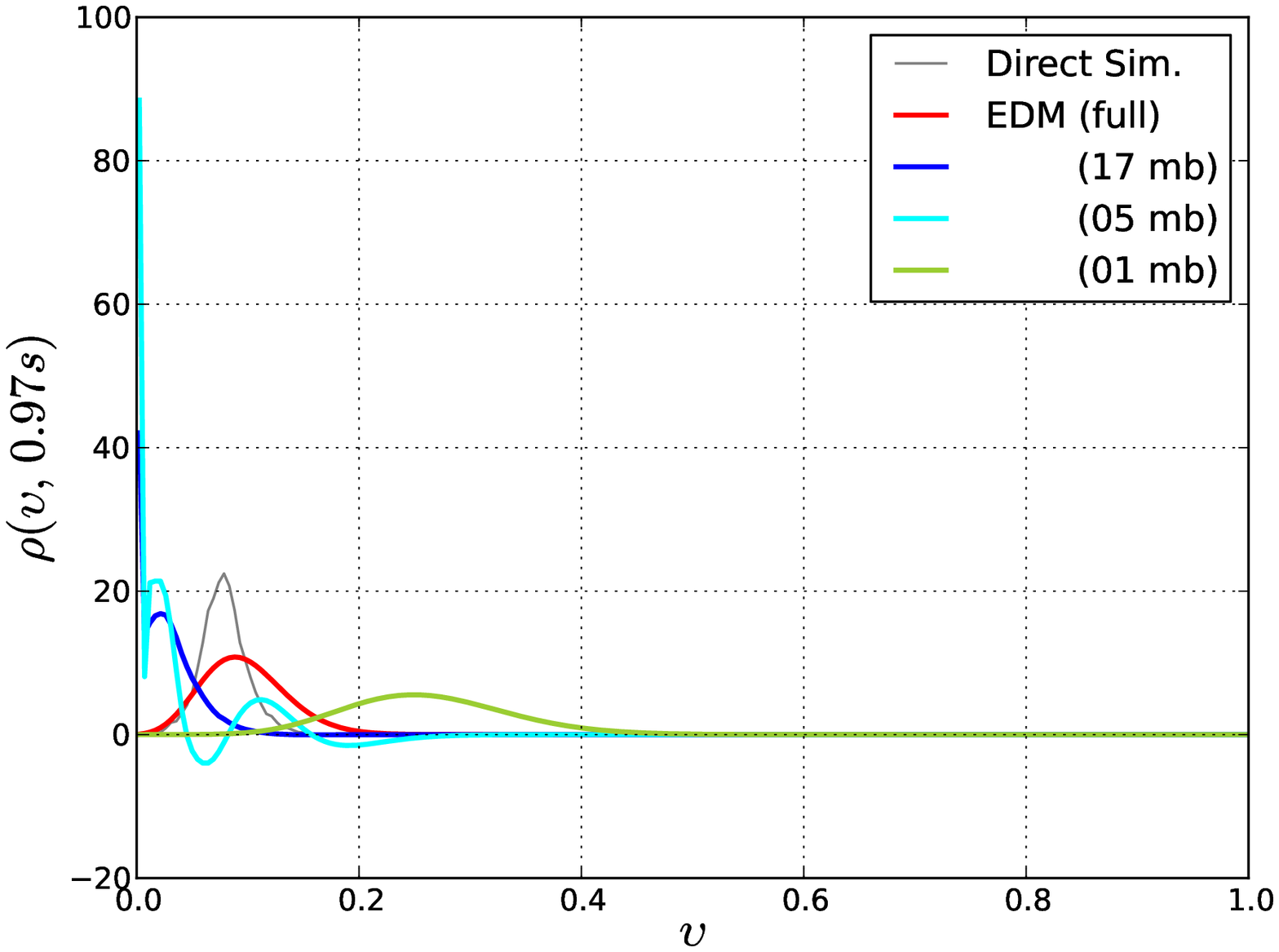}
\end{center}

\caption{Ensemble pdfs for the inhibitory ensemble 0.97s.}

\label{fig:iRhosAt0.97}
\end{figure}

\begin{figure}
\begin{center}
\includegraphics[width=5in]{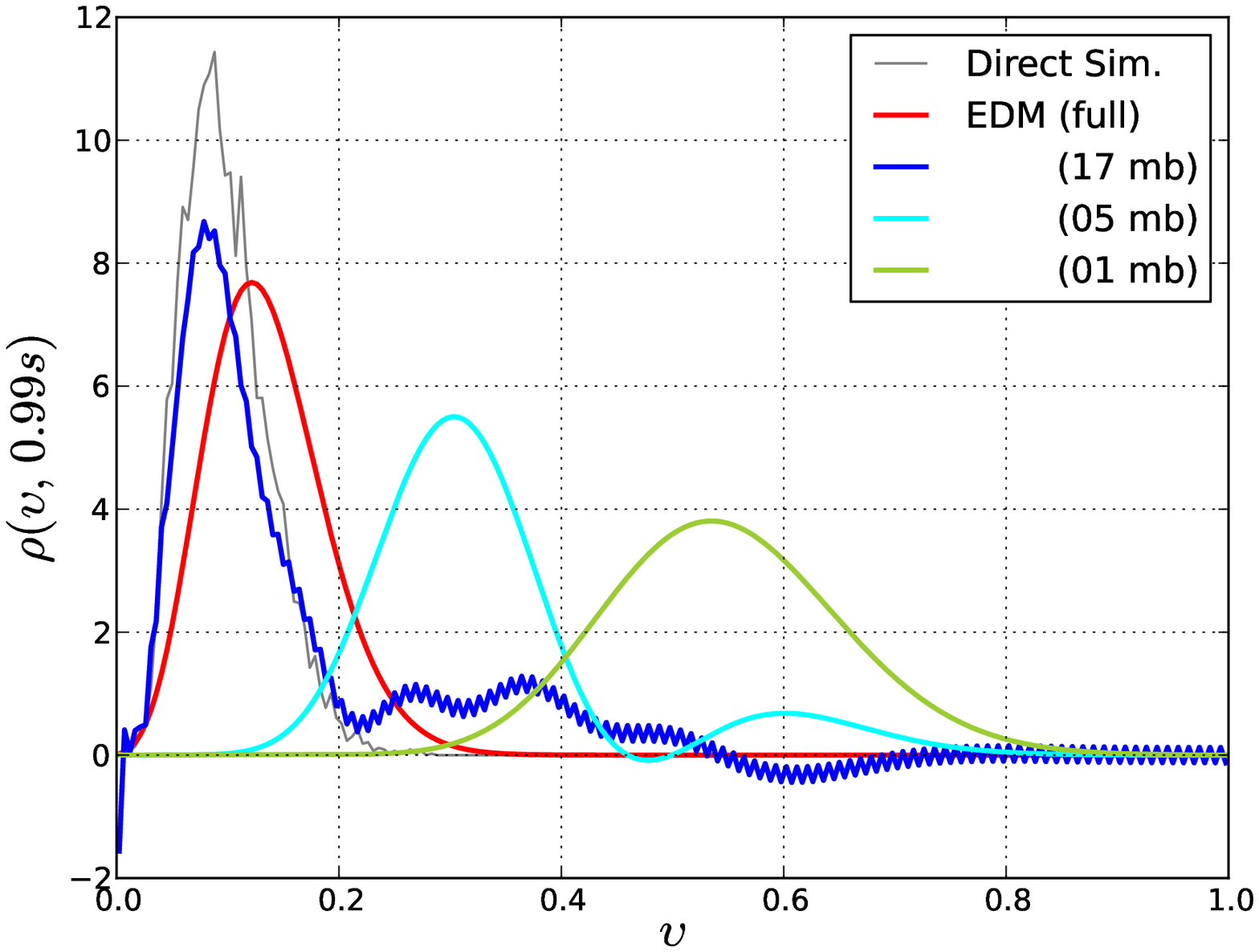}
\end{center}

\caption{Ensemble pdfs for the inhibitory ensemble 0.99s.}

\label{fig:iRhosAt0.99}
\end{figure}

\begin{figure}
\begin{center}
\includegraphics[width=5in]{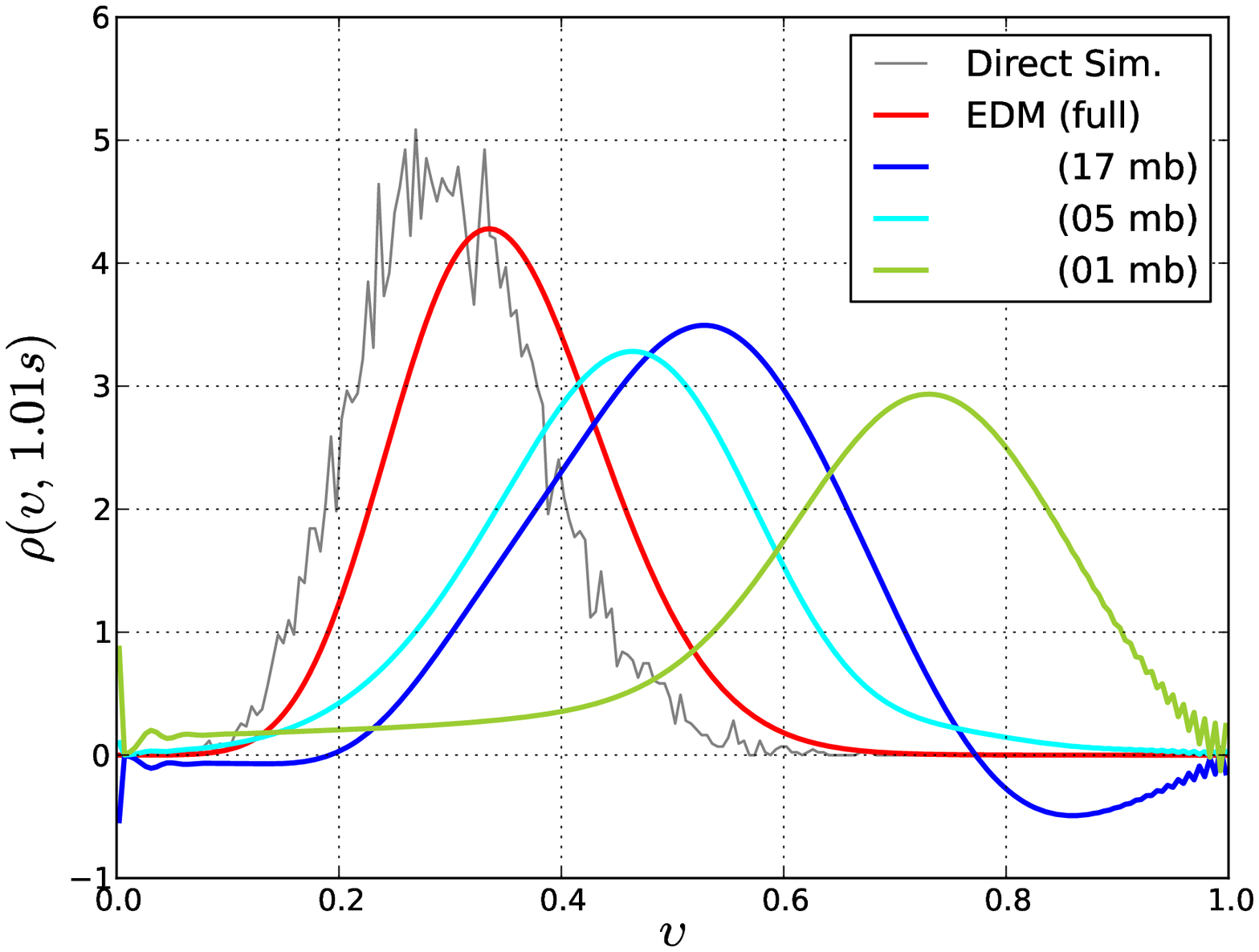}
\end{center}

\caption{Ensemble pdfs for the inhibitory ensemble 1.01s.}

\label{fig:iRhosAt1.01}
\end{figure}

\begin{figure}
\begin{center}
\includegraphics[width=5in]{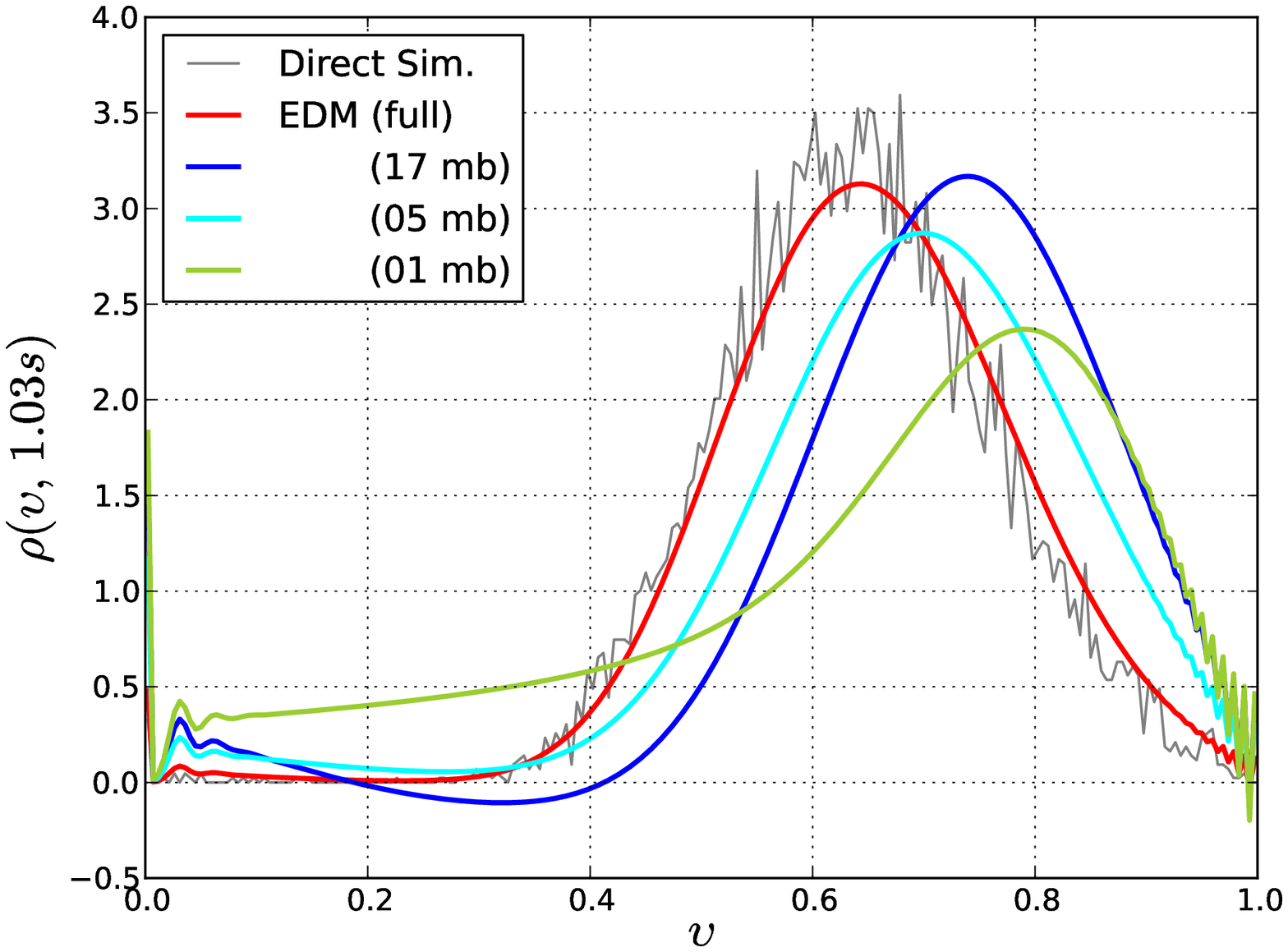}
\end{center}

\caption{Ensemble pdfs for the inhibitory ensemble 1.03s.}

\label{fig:iRhosAt1.03}
\end{figure}

\begin{figure}
\begin{center}
\includegraphics[width=5in]{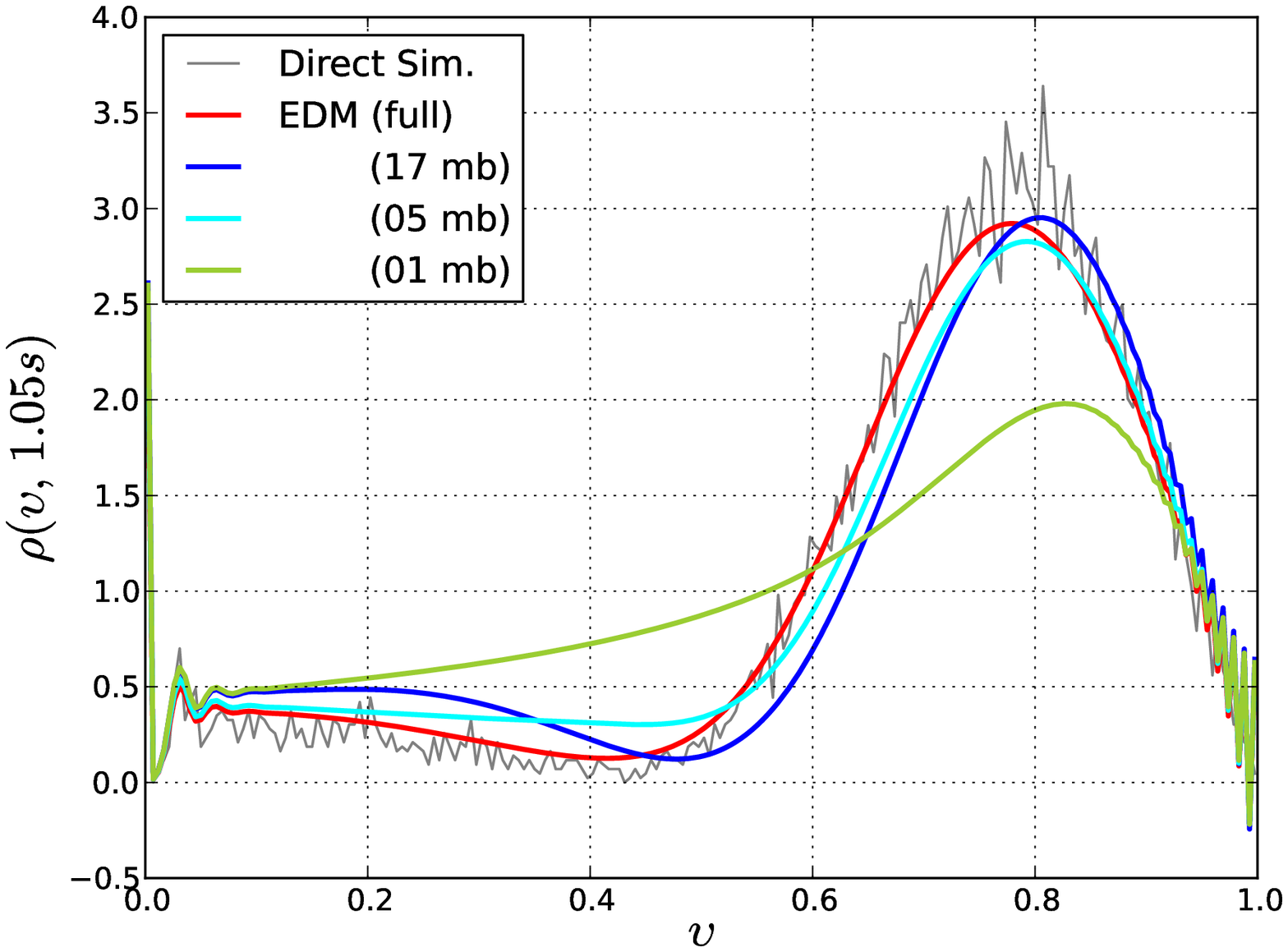}
\end{center}

\caption{Ensemble pdfs for the inhibitory ensemble 1.05s.}

\label{fig:iRhosAt1.05}
\end{figure}

\begin{figure}
\begin{center}
\includegraphics[width=5in]{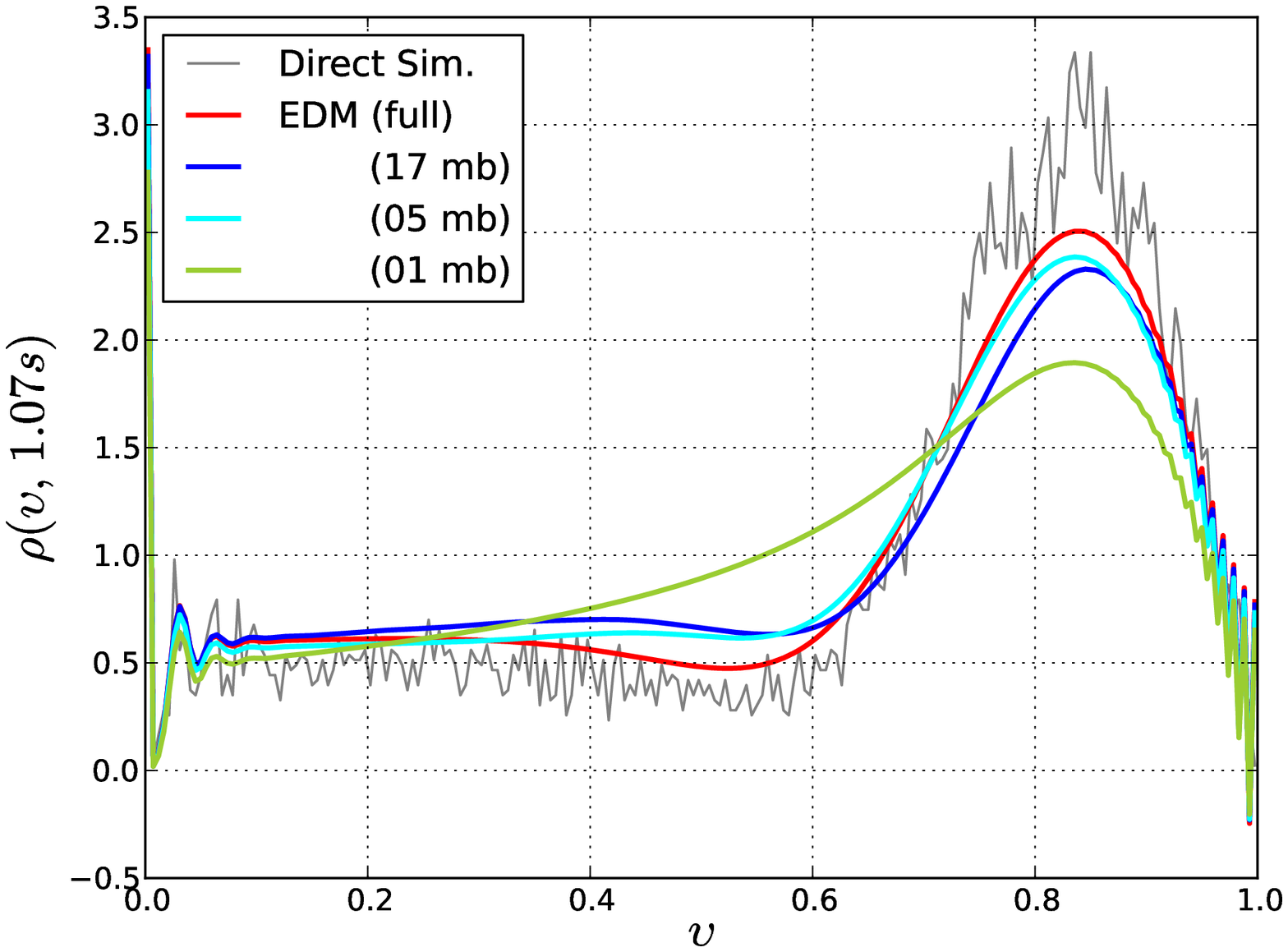}
\end{center}

\caption{Ensemble pdfs for the inhibitory ensemble 1.07s.}

\label{fig:iRhosAt1.07}
\end{figure}

\subsection{Partial conclusions}

We have developed a method to reduce the dimensionality of the state space of
EDMs.  We observed that the pdfs from low-dimensional EDMs move faster to and
away from the high-average-firing-rate-per-neuron pdf than the pdf derived from
direct simulation or that from full EDMS.  However, these differences did not
have a large impact on the average firing rate per neuron produced by
low-dimensional EDMs, which almost perfectly approximates the average firing
rate per neuron computed by direct simulation.

Our next step is to estimate connectivity parameters and state space
variables in networks of EDMs from recorded spike rates. We will first do
this with simulated data.

\pagebreak

\section{Estimating parameters of EDMs}
\label{sec:estimatingParamsOfEDMs}

Here we evaluate a maximum likelihood method to estimate the connectivity
parameters $\bm{\theta}=[W_{ei},W_{ie}]$ in the EDMs network in
Figure~\ref{fig:edmsNetwork}.  We seek the connectivity parameters for which a
set of $N$ pairs measurements,  $Y_N=[\mathbf{y}(0), \ldots, \mathbf{y}(N-1)]$,
are most probable; ie:

\begin{figure}
\begin{center}
\includegraphics[width=4in]{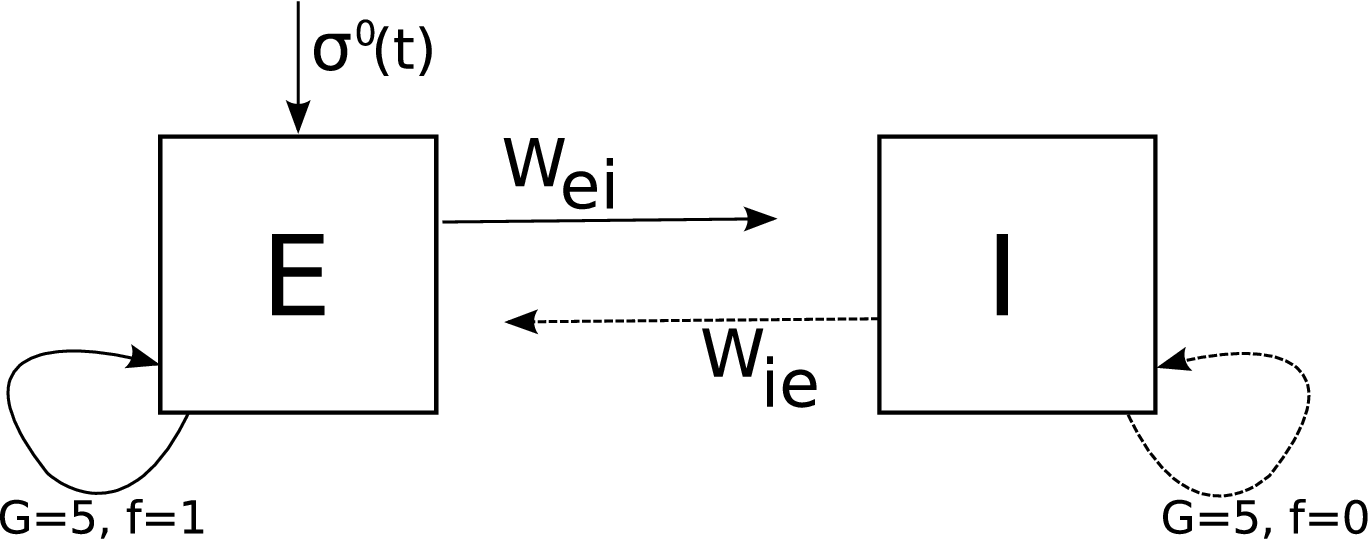}
\end{center}

\caption{Network of EDMs.}

\label{fig:edmsNetwork}
\end{figure}

\begin{eqnarray*}
\bm{\theta}_{ml}=\argmax_{\bm{\theta}}P(Y_N|\bm{\theta})
\end{eqnarray*}

\noindent The pair of measurements at time $n$, $\mathbf{y}(n)$ comprises
measurements from the excitatory and inhibitory ensembles; i.e.,
$\mathbf{y}(n)=[y_e(n), y_i(n)]$.

\subsection{Noise Model}

As a first approximation we make the following three assumptions on the noise
of the measurements. These are the assumptions made in previous work by our
collaborators \citep{kostukEtAl12}.

\begin{enumerate}

\item Noise is independent in time; i.e.,
$P(Y_N|\bm{\theta})=\Pi_{n=1}^NP(\mathbf{y}(n)|\bm{\theta})$. 

\item Noise is independent in the different populations; i.e.,
$P(\mathbf{y}(n)|\bm{\theta})=P(y_e(n)|\bm{\theta})P(y_i(n)|\bm{\theta})$.

\item Gaussian noise with a known variance, $\sigma^2$, in each population;
i.e., $P(y_.(n)|\bm{\theta})=N(y_.(n)|r_.(n|\bm{\theta}), \sigma^2)$, where
$r_.(n|\bm{\theta})$ is the activity generated by the ensemble at time $n$.  

\end{enumerate}

With these assumptions the log likelihood function of
the model parameters reduces to:

\begin{eqnarray*}
\log P(Y_N|\bm{\theta})=K-\Sigma_{n=0}^N\frac{(y_e(n)-r_e(n|\bm{\theta}))^2+(y_e(n)-r_e(n|\bm{\theta}))^2}{2\sigma^2}
\end{eqnarray*}

\begin{figure}
\begin{center}
\includegraphics[width=6in]{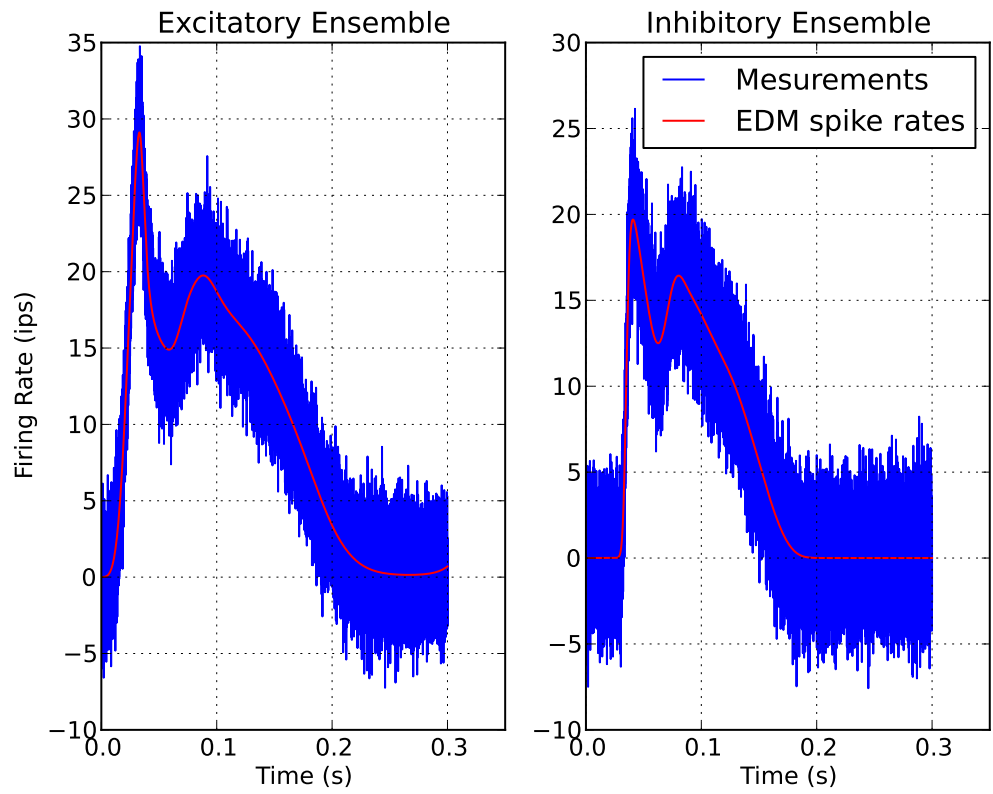}
\end{center}

\caption{EDMs spike rates (red) and noisy measurements (blue, $\sigma=2$) used
to estimate the connectivity parameters in the network of
Figure~\ref{fig:edmsNetwork}.}

\label{fig:ys}
\end{figure}

The red lines in Figure~\ref{fig:ys} plots the activity generated by the
excitatory and inhibitory ensemble with connectivity parameters $W_{ei}=50$ and
$W_{ie}=15$. The blue lines plot the noisy measurements ($\sigma=2$) that we
use below for parameter estimation.

\subsection{Optimization Surface}

\begin{figure}
\begin{center}
\includegraphics[width=6in]{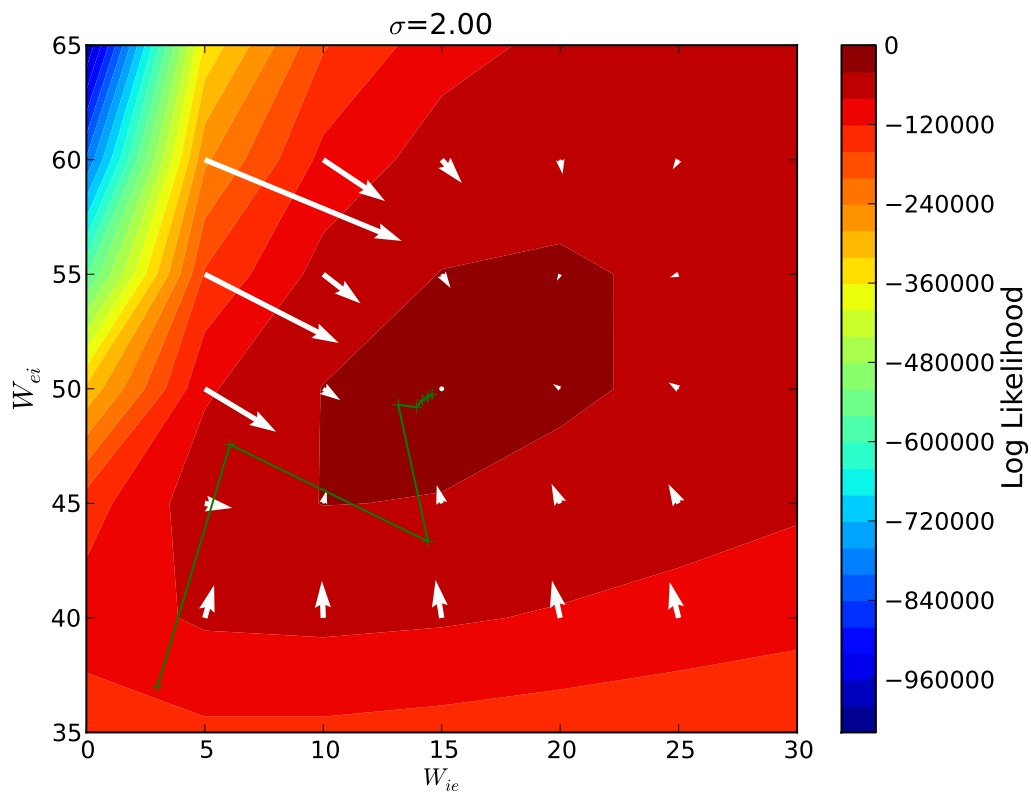}
\end{center}

\caption{Log-likelihood function, its gradient, and a sample gradient ascent
path. The contour plot plots the log-likelihood function for the parameter
values shown on the axes for the network in Figure~\ref{fig:edmsNetwork}. The white arrows show the log-likelihood gradient
computed analytically. The green curve is a gradient ascent trajectory starting
at $W_{ei}=37$ and $W_{ie}=3$.}

\label{fig:optimization}
\end{figure}

The levelplot in Figure~\ref{fig:optimization} shows the optimization surface
for the set of parameters show in the axes. We see that it has a convex shape
with peak at the true parameters. Thus, an iterative gradient-ascent
optimization procedure should climb to the maximum-likelihood parameters from
any starting set of parameters.

\subsection{Gradient of Log-Likelihood Function}

To compute the gradient of the log-likelihood function at a given set of
parameters $\bm{\theta}=[w_{ei}$, $w_{ie}]$ one needs to integrate the EDMs to
generate at each time step the activities of both ensembles,
$r_e(n,\bm{\theta})$ and $r_i(n,\bm{\theta})$, as well as the ensembles pdfs,
$\bm{\rho}_e(n|\bm{\theta})$ and $\bm{\rho}_i(n|\bm{\theta})$. With these
quantities in hand, the gradient of the log-likelihood function can be computed
recursively in a second integration step (Equation~\ref{eq:llGradient}).

\begin{eqnarray}
    \frac{\partial\log P(Y_m|\bm{\theta})}{\partial W_{ei}}&=&\Sigma_{n=0}^m\frac{(y_e[n]-r_e[n;\bm{\theta}])\frac{\partial r_e[n;\bm{\theta}]}{\partial W_{ei}}+(y_i[n]-r_i[n;\bm{\theta}])\frac{\partial r_i[n;\bm{\theta}]}{\partial W_{ei}}}{\sigma^2}\nonumber\\
    \frac{\partial r_e[n;\bm{\theta}]}{\partial W_{ei}}&=&\Delta v\,\sigma_e^E[n]\, (\mathbf{q}_r,\frac{\partial\bm{\rho_e}[n|\bm{\theta}]}{\partial W_{ei}})\nonumber\\
    \frac{\partial\bm{\rho}_e[n|\bm{\theta}]}{\partial W_{ei}}&=&\Delta t \frac{\partial Q_e[n-1;\bm{\theta}]}{\partial W_{ei}}\bm{\rho}_e[n-1|\bm{\theta}]+\left[I+\Delta t \, Q_e[n-1;\bm{\theta}]\right]\frac{\partial\bm{\rho}_e[n-1|\bm{\theta}]}{\partial W_{ei}}\nonumber\\
    \frac{\partial Q_e[n;\bm{\theta}]}{\partial W_{ei}}&=&-W_{ei}\frac{\partial r_i[n-1;\bm{\theta}]}{\partial W_{ei}}A^{(2)}
\label{eq:llGradient}
\end{eqnarray}

The white arrows in Figure~\ref{fig:optimization} point in the direction of the
gradient, and are scaled according to the gradient magnitude. Note that, as
expected, the gradient direction is perpendicular to the level lines.

With a convex log-likelihood function, for which we can compute its gradient,
we can now use an iterative gradient ascent procedure to maximize this
function. The green line in Figure~\ref{fig:optimization} shows a
gradient-ascent trajectory that in only three steps accurately approximated the
maximum likelihood parameters.

\pagebreak

\section{Modeling evoked auditory activity in rodents with EDMs}
\label{sec:modelingRodentsRecordingsWithEDMs}

Here we show that the average firing-rate per neuron predicted by EDMs well
approximates the evoked high-gamma power (HGP) from auditory neurons in
response to stimulation with pure-tones.

\subsection{Recordings}

\begin{figure}
\begin{center}
    \includegraphics[width=5in]{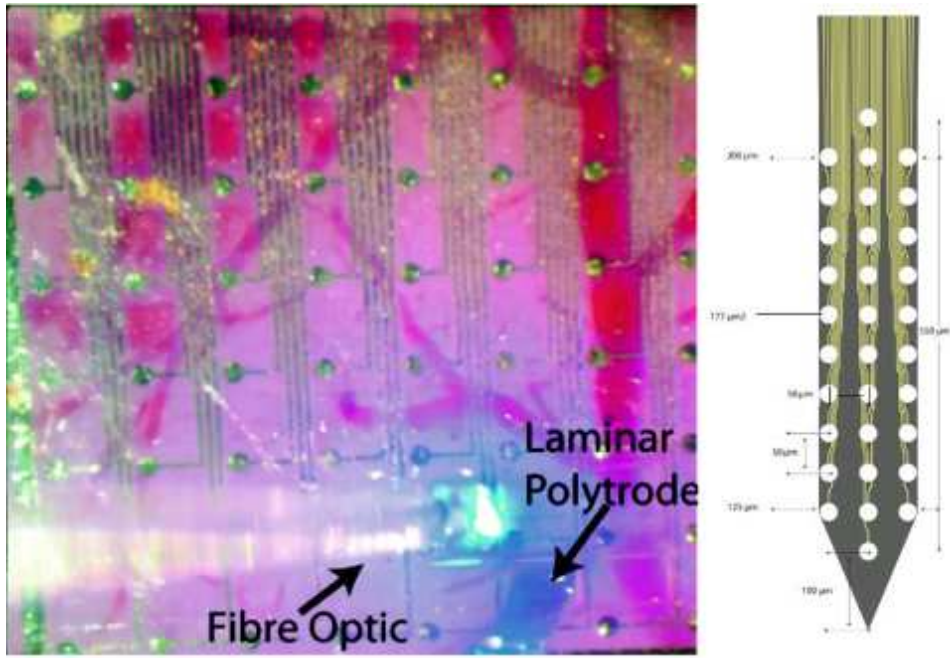}
\end{center}

\caption{Electrodes used for simultaneous surface and laminar recordings.
Surface recordings were obtained from an $8\times8$ grid of subdurally
implanted electrodes covering an area of $1.6~mm^2$. Laminar recordings were
obtained from a 32-channel politrodes of length $650~\mu m$.  These rercordings
were combined with optogenetic manipulations.}
\label{fig:recordignElectrodes}

\end{figure}

\begin{figure}
\begin{center}
    \begin{center}
        \includegraphics[width=5in]{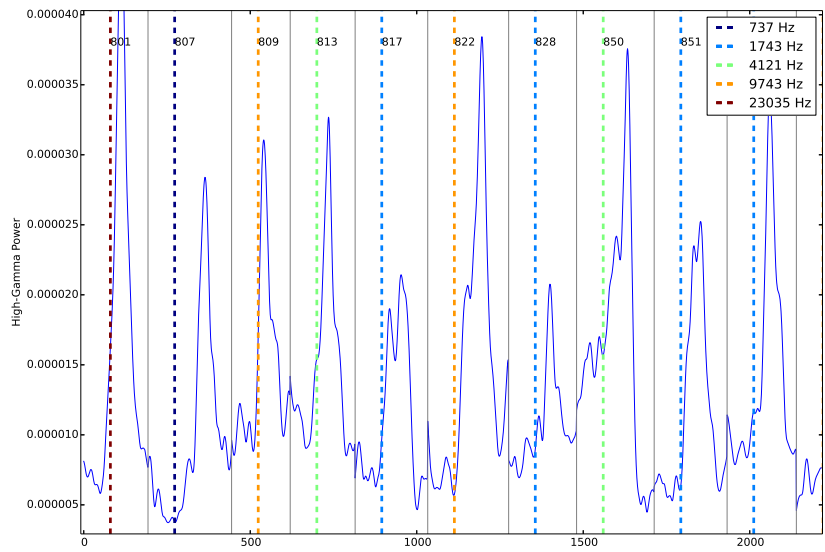}
    \end{center}
\end{center}

\caption{
High-gamma power evoked by the presentation of pure tones at different
frequencies and amplitudes. Blue traces show the HGP evoked by the
presentation of tones in one sample electrode of the array. Each vertical dotted
color line marks the onset of a tone of a corresponding frequency.
}

\label{fig:hgp1}

\end{figure}

We used very high-resolution simultaneous surface and laminar recordings
(Figure~\ref{fig:recordignElectrodes}) in anesthetized rodents stimulated with
pure tones at different frequencies and amplitudes.  The blue trace in
Figure~\ref{fig:hgp1} shows the high-gamma power (HGP) evoked by the
presentation of tones in one sample electrode of the array. Each vertical
dotted color line marks the onset of a tone of a corresponding frequency.

\begin{figure}
\begin{center}
    \begin{center}
        \includegraphics[width=5in]{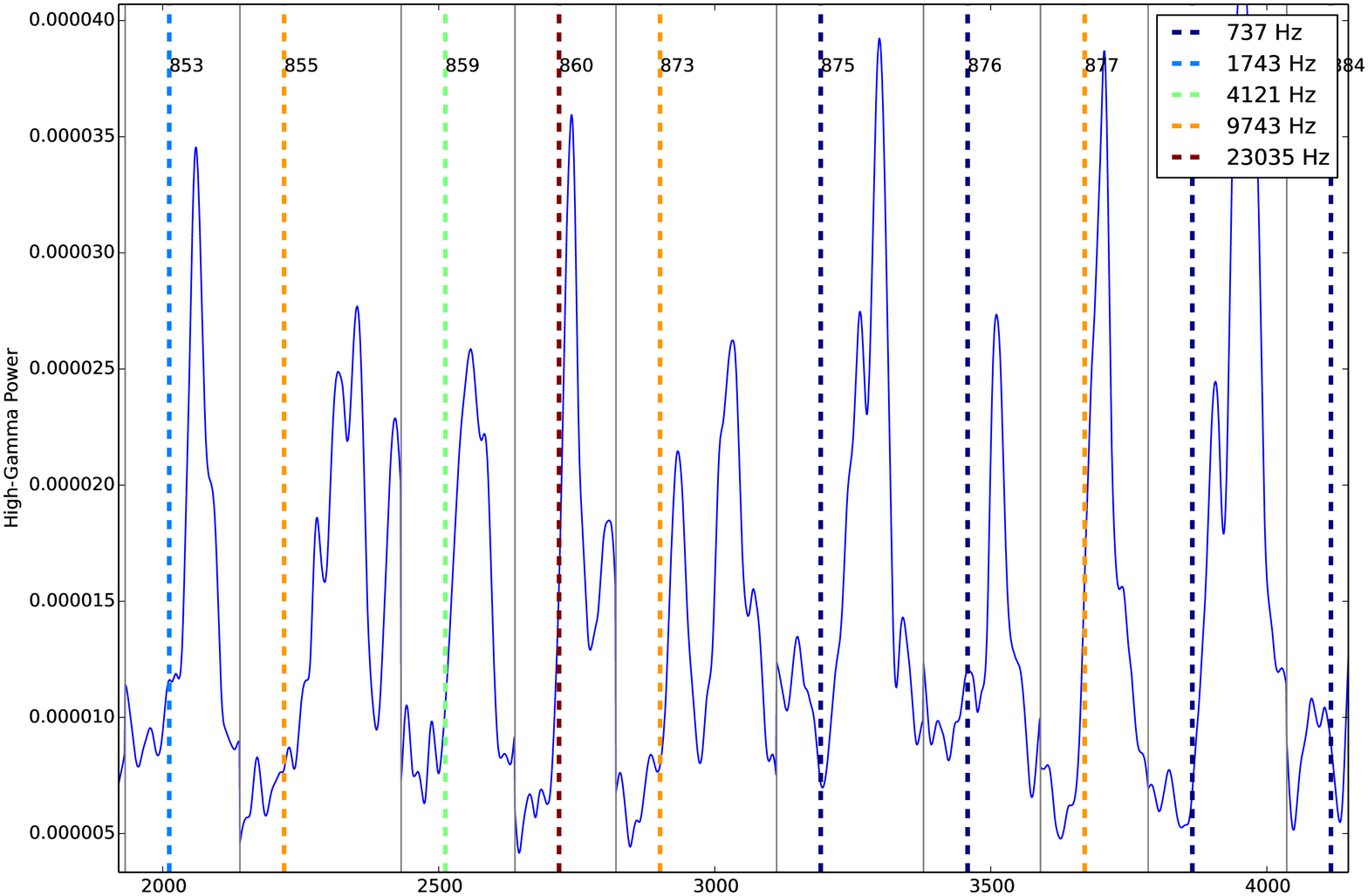}
    \end{center}
\end{center}

\caption{
More examples of HGP evoked by the presentation of pure tones. The
format is as in Figure~\ref{fig:hgp1}.
}

\label{fig:hgp2}

\end{figure}

We see that these evoked waveforms are very stereotypical. Some waveforms have
a large peak followed by a bump. Other waveforms have two peaks, with a smaller
peak preceding a larger one.  We also see inf Figure~\ref{fig:hgp2} waveforms
that have more than one peak preceding a larger one and waveforms having a
larger peak followed by a smaller one.

\subsection{Qualitative analysis}

\begin{figure}
\begin{center}
    \includegraphics[width=5in]{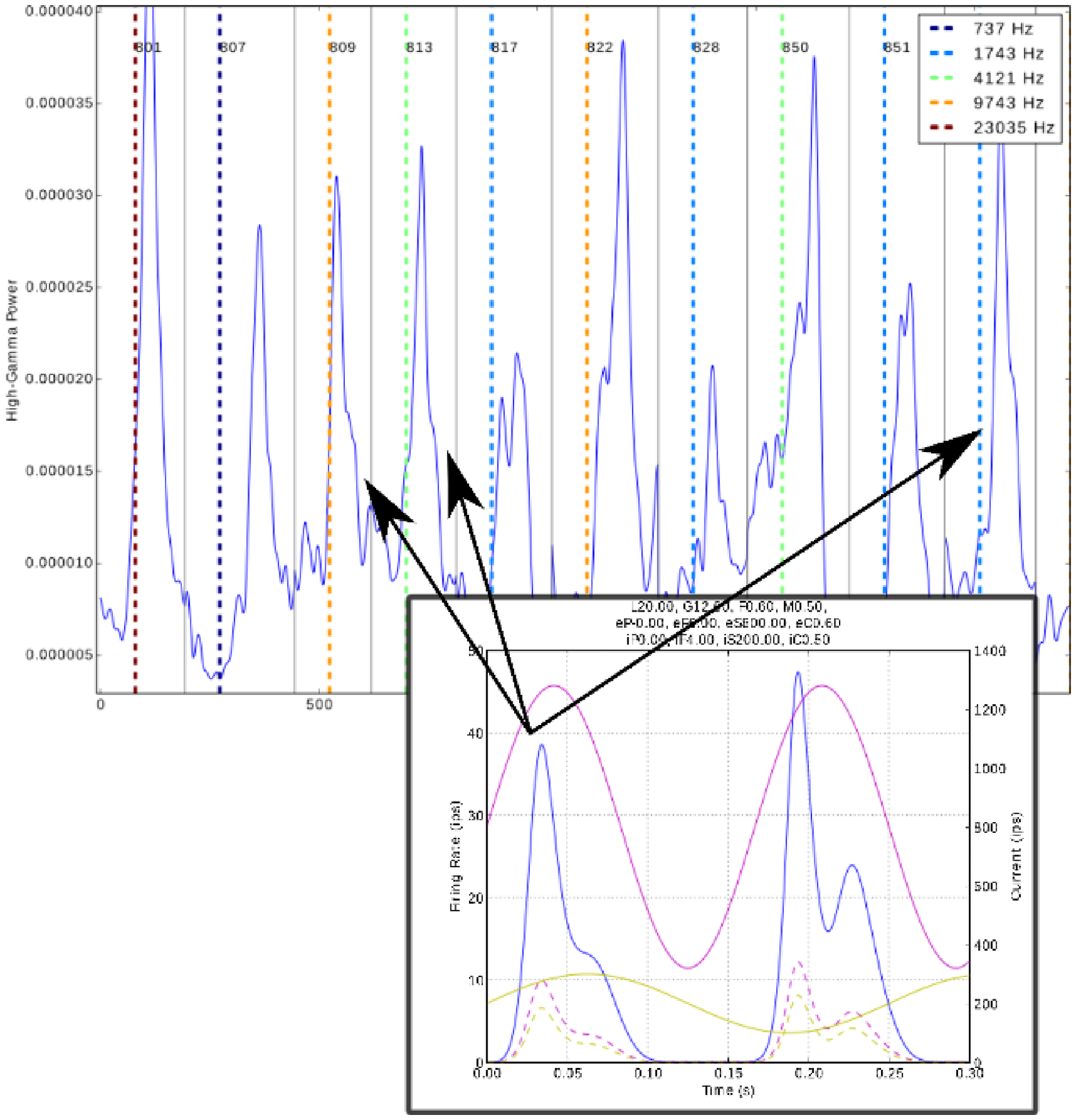}
\end{center}

\caption{The blue curve in the inset shows the average firing-rate simulated by
an EDM with manually chosen parameters. The magenta and yellow solid curves
plot its excitatory and inhibitory inputs, respectively.  The first waveform is
qualitatively similar to the ECoG waveforms with a bump following a large
peak.}
\label{fig:manualFit1}

\end{figure}

\begin{figure}
\begin{center}
    \includegraphics[width=5in]{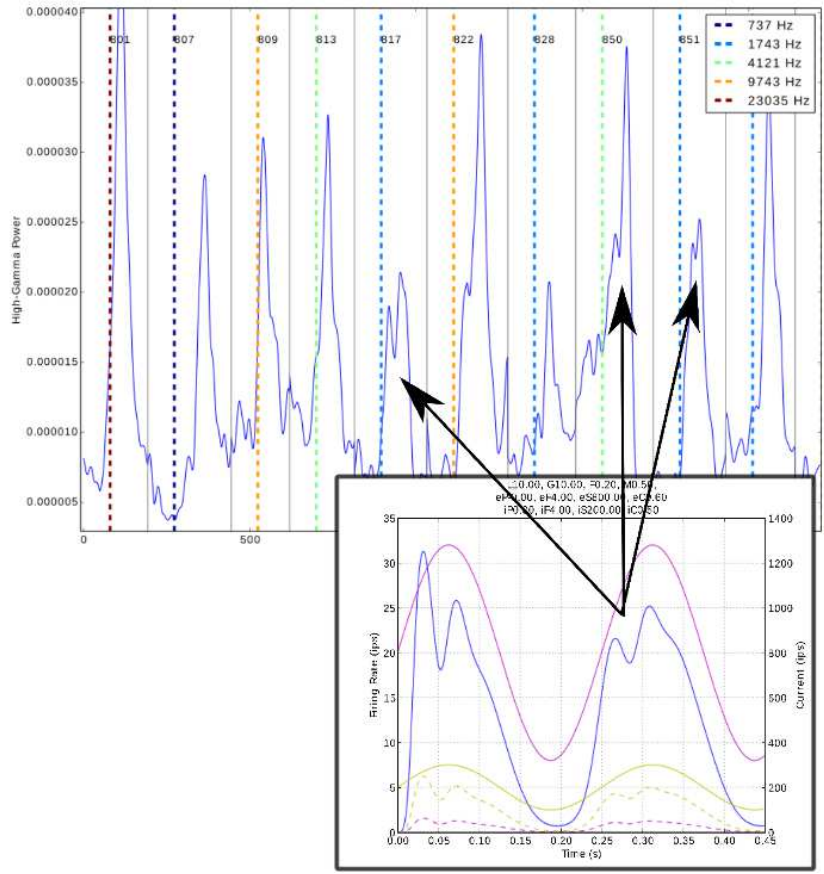}
\end{center}

\caption{The blue curve in the inset shows the average firing-rate simulated by
an EDM with manually chosen parameters. The magenta and yellow solid curves
plot its excitatory and inhibitory inputs, respectively.  The second waveform
is qualitatively similar to the ECoG waveforms with a lower peak preceding a
large one.}

\label{fig:manualFit2}

\end{figure}

\begin{figure}
\begin{center}
    \includegraphics[width=5in]{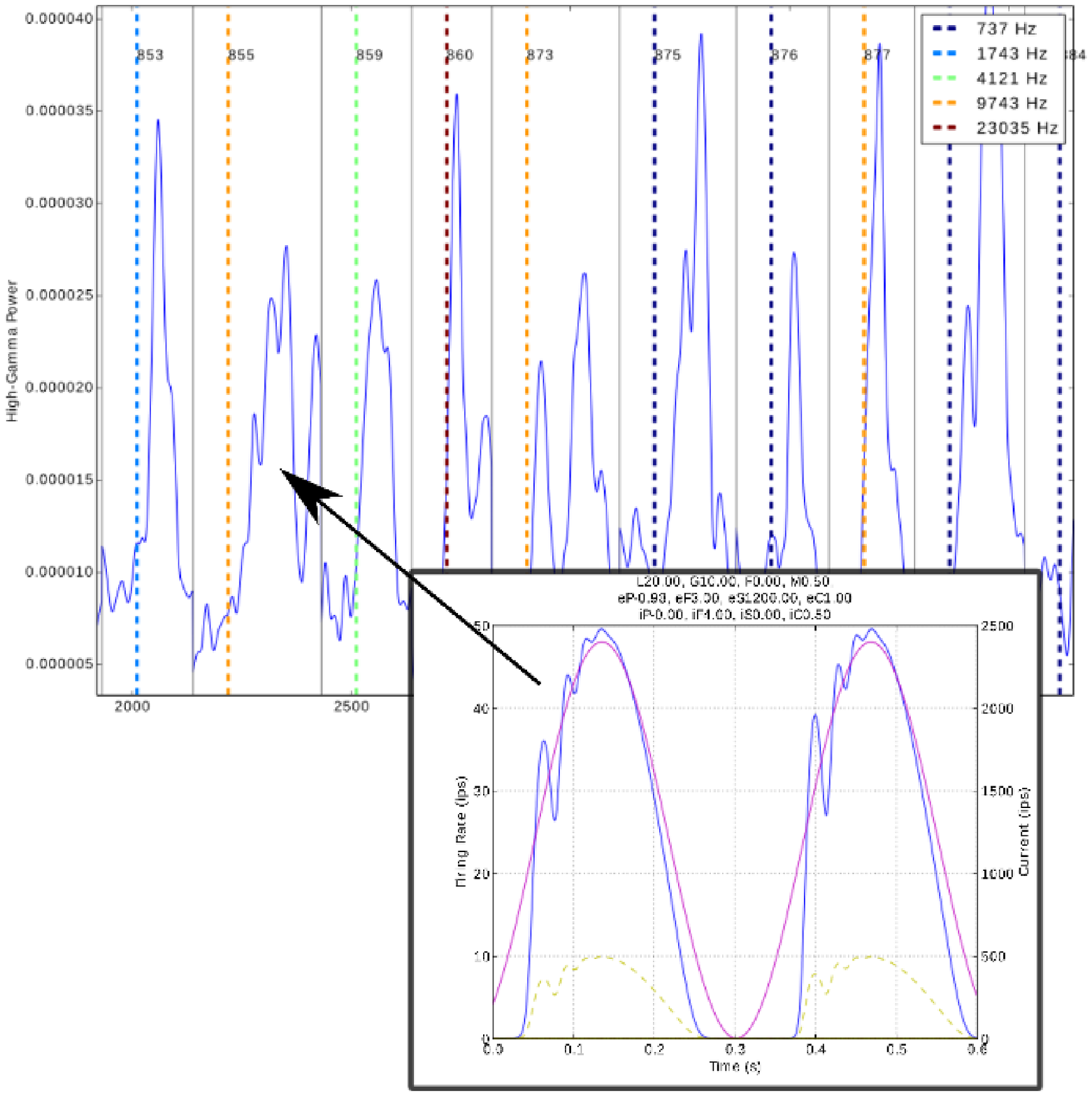}
\end{center}

\caption{The blue curve in the inset shows the average firing-rate simulated by
an EDM with manually chosen parameters. The magenta and yellow solid curves
plot its excitatory and inhibitory inputs, respectively. EDMs can generate
waveforms with multiple lower peaks preceding a larger one.}

\label{fig:manualFit3}

\end{figure}

\begin{figure}
\begin{center}
    \includegraphics[width=5in]{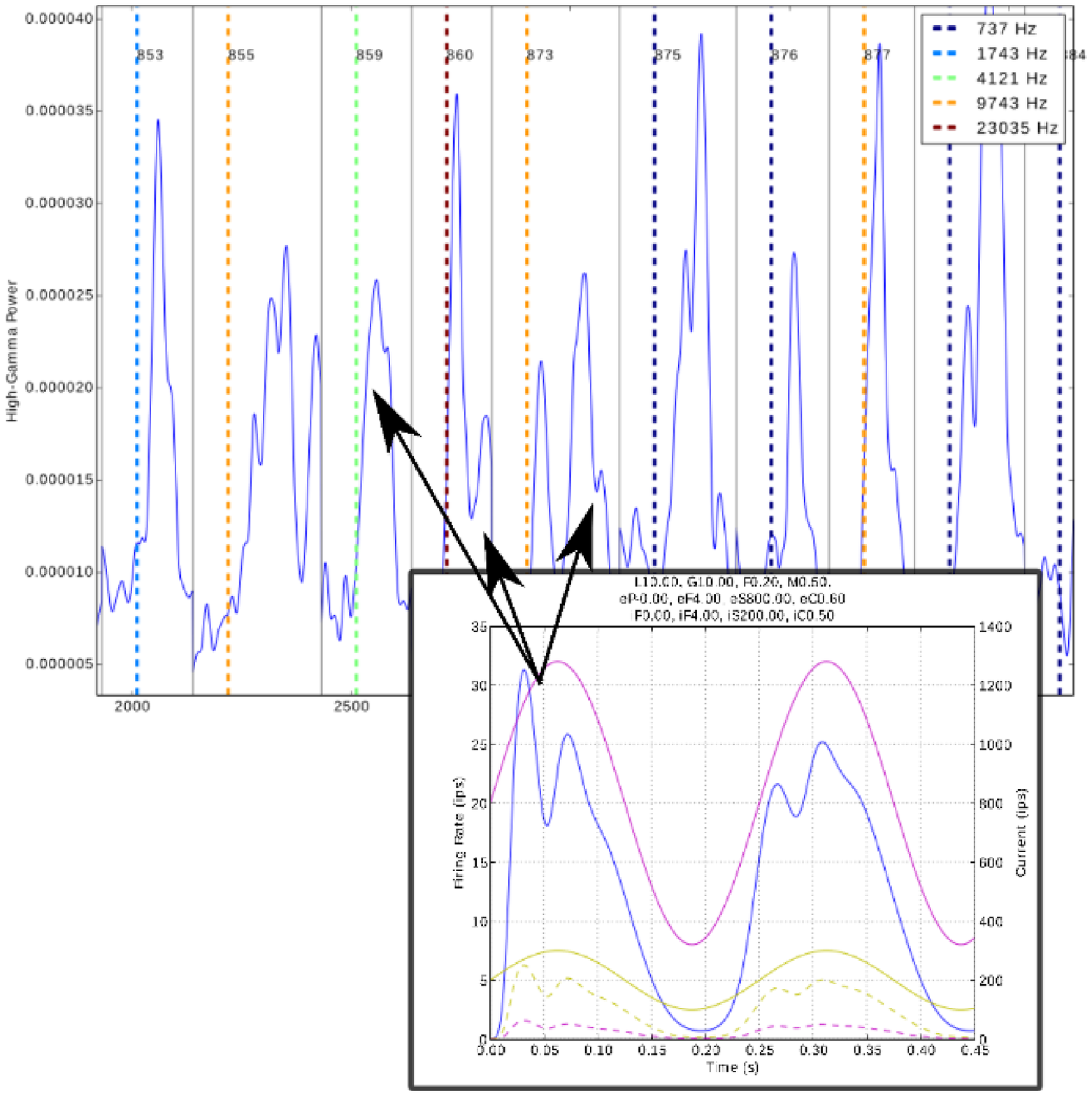}
\end{center}

\caption{The blue curve in the inset shows the average firing-rate simulated by
an EDM with manually chosen parameters. The magenta and yellow solid curves
plot its excitatory and inhibitory inputs, respectively. EDMs can generate
waveforms with a larger peak preceding a smaller one.}

\label{fig:manualFit4}

\end{figure}

The recorded HGP waveforms appear remarkably similar to those produced by EDMs
when stimulated by sinusoids. To confirm this similarity we manually chose
parameters for EDMs to reproduce the shape of recorded waveforms
(Figures~\ref{fig:manualFit1}, \ref{fig:manualFit2}, \ref{fig:manualFit3}, and
\ref{fig:manualFit4}).

\subsection{Quantitative analysis}

\begin{figure}
\begin{center}
    \includegraphics[width=5in]{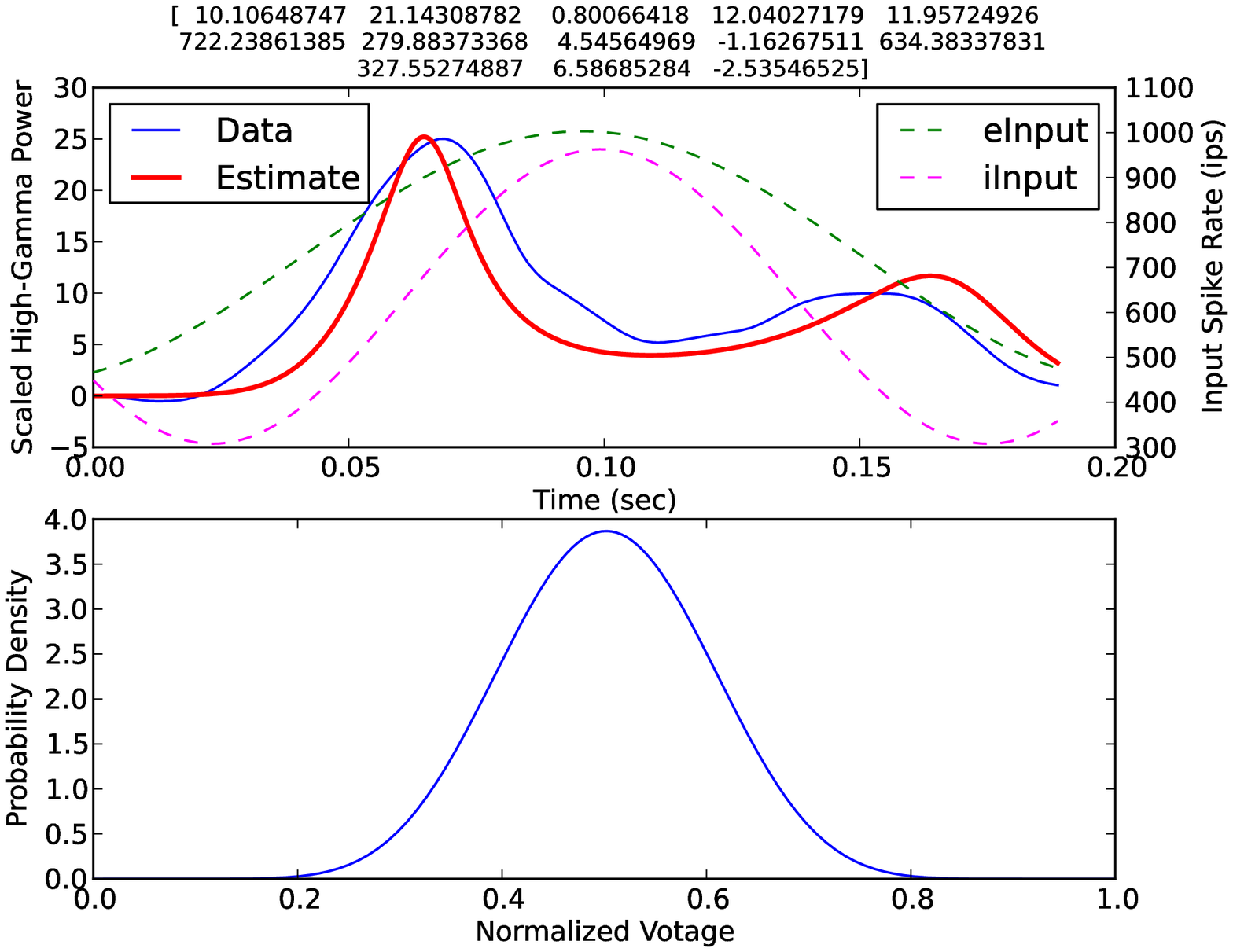}
\end{center}

\caption{Learning ensemble properties to approximate physiological recordings.
We estimated 13 parameters so that the average firing rate per neuron predicted
by an EDM approximates as close as possible the recorded HGP in response of a
pure tone.  The title shows the values of these parameters. The blue and red
curves in the top panel plot the recorded HGP and the predicted average firing
rate per neuron, respectively. The dotted green and magenta curves in the top
panel shows the estimated excitatory and inhibitory inputs to the EDM. The
curve in the bottom panel plot the estimated initial condition probability
density function of the EDM. The average firing rate predicted by the EDM is a
good approximation of the recorded HGP.}

\label{fig:automaticFit}

\end{figure}

Using a similar method as described above to learn connectivity parameters in
networks of EDMs, we etimated parameters so that an EDM approximated as close
as possible the HGP evoked by the presentation of pure tones to rodents. In
total we estimated 13 parameters: three parameters of an EDM, two parameters
for its initial condition, four parameters for the sinusoidal excitatory input,
and four parameters for the inhibitory input.  The recorded and approximated
HGP are given by the blue and red solid curves, respectively, in
Figure~\ref{fig:automaticFit}.

\subsection{Partial Conclusions}

These preliminary results show that EDMs, besides accurately approximating the
average firing rate of ensembles of simulated IF neurons, well approximate the
HGP in the auditory cortex of rats evoked by auditory tones.

\section{Conclusions}

We have shown that an ensemble model accurately reproduce the probability
density function of the transmembrane voltage, as well as the average-firing
rate per neurona, in a large ensemble of integrate-and-fire simulated
neurons (Section~\ref{sec:buildingEDMs}). We developed and evaluated methods to
reduce the dimensionality (Section~\ref{sec:reducingDimOfEDMs}) and estimate
parameters (Section~\ref{sec:estimatingParamsOfEDMs}). Finally we demonstrated
the feasibility of EDMs to model the high-gamma power evoked by pure tones in
the auditory cortex of rodents.

The possiblity of quantitatively model the activity of ensemble of neurons may
allow us to uncover fundamental computations performed by neural ensembles.

\bibliographystyle{plainnat}
\bibliography{mesoscaleModels,estimationOfDynamicalSystems}

\end{document}